\documentclass[english,prd,nofootinbib,preprintnumbers,twocolumn,showpacs]{revtex4}
\usepackage[latin1]{inputenc}
\usepackage{amsmath}
\usepackage{amsfonts}
\usepackage{appendix}
\usepackage{amssymb}
\usepackage{epsfig}
\usepackage{graphics,psfrag,rotating}
\usepackage{graphicx}
\usepackage{dcolumn}
\usepackage{bm}
\usepackage{epstopdf}
\usepackage{color}
\usepackage[usenames,dvipsnames,svgnames]{xcolor}
\usepackage[colorlinks=true,
            linkcolor=red,
            urlcolor=gray,
            citecolor=blue]{hyperref}
\usepackage{hyperref}
\usepackage[T1]{fontenc}
\usepackage{multirow}
\usepackage{float}
\usepackage{cancel}
\usepackage{hyperref}
\usepackage{adjustbox}
\usepackage{subfigure}

\usepackage{enumitem}

\def\3nab{\tilde{\nabla}}

\def\be {\begin{equation}}
\def\ee {\end{equation}}
\def\ba {\begin{align}}
\def\ea {\end{align}}

\def\bc {\begin{center}}
\def\ec {\end{center}}
\def\case#1/#2{\frac{#1}{#2}}

\newcommand{\bver}{\begin{verbatim}}
\newcommand{\ever}{\end{verbatim}}

\newcommand{\bea}{\begin{eqnarray}}
\newcommand{\eea}{\end{eqnarray}}
\newcommand{\beaa}{\begin{eqnarray*}}
\newcommand{\eeaa}{\end{eqnarray*}}

\def\case#1/#2{\textstyle\frac{#1}{#2}}

\begin{document}

%%%%%%%%%%%%%%%%%%%%%%%%%%%%%%%%%%%%%%%%%%%%%%%%%%%%%%%%%%%
\title{Testing for gravitational preferred directions with galaxy and lensing surveys}
%%%%%%%%%%%%%%%%%%%%%%%%%%%%%%%%%%%%%%%%%%%%%%%%%%%%%%%%%%%

\author{
Miguel Aparicio Resco 
}
\email{migueapa@ucm.es}
\affiliation{Departamento de F\'{\i}sica Te\'orica and Instituto de F\'{\i}sica de Part\'{\i}culas y del Cosmos IPARCOS, Universidad Complutense de Madrid, 28040 
Madrid, Spain}
\author{Antonio L.\ Maroto}
\email{maroto@ucm.es}
\affiliation{Departamento de F\'{\i}sica Te\'orica and Instituto de F\'{\i}sica de Part\'{\i}culas y del Cosmos IPARCOS, Universidad Complutense de Madrid, 28040 
Madrid, Spain}


%%%%%%%%%%%%%

%\date{\today}

\begin{abstract} 

We analyze the sensitivity of galaxy and weak-lensing surveys to detect preferred
directions  in the gravitational interaction. 
We consider general theories of gravity  involving additional vector degrees of freedom with non-vanishing spatial components in the background. We use a model-independent parametrization of the perturbations equations in terms of four effective parameters, namely, the standard effective Newton constant $G_{eff}$ and slip parameter $\gamma$  for scalar modes and two new parameters $\mu_Q$ and $\mu_h$ for vector and tensor modes respectively, which are required when preferred directions are present. We obtain the 
expressions for the multipole galaxy power spectrum 
in redshift space and for the weak-lensing shear, convergence and rotation spectra in the presence of preferred directions. By performing a Fisher matrix forecast analysis, we estimate the sensitivity of a future Euclid-like survey to detect this kind of modification of gravity.
We finally compare with the effects induced by violations of statistical isotropy in the primordial power spectrum and identify the observables which could discriminate
between them.

\end{abstract} 

\maketitle

%%%%%%%%%%%%%%%%%%%%%%%%%%%%%%%%%%%%%%%%%%%%%%%%%%%%%%%%%%%%%%%%%%%%%%%%%%%%%%%%%%%%
\section{Introduction}
%%%%%%%%%%%%%%%%%%%%%%%%%%%%%%%%%%%%%%%%%%%%%%%%%%%%%%%%%%%%%%%%%%%%%%%%%%%%%%%%%%%%
Rotational invariance, as part of the Lorentz group, is one of the underlying symmetries in our current  description of the fundamental interactions of nature.
The weak Equivalence Principle, which is one of the cornerstones of General Relativity (GR), ensures that Lorentz invariance is  respected not only in flat space-time, but also in the presence of gravity, where  the symmetry is locally preserved \cite{Weinberg,Will,Shao:2016ezh}. 

On the other hand, it is also well established from current observations \cite{Ade:2015hxq} that rotational symmetry is also manifested in a statistical 
way on the large-scale distribution of matter and radiation in the universe. In the standard
inflationary scenario, density perturbations are generated from quantum 
vacuum fluctuations, so that the isotropy of the primordial spectrum of perturbation reflects the invariance under rotations of the
quantum vacuum state \cite{Lyth,Ackerman:2007nb}.

Despite the fact that our current description of interactions  seems to be 
compatible with rotational invariance on a wide range
of scales, certain observations seem to suggest
the existence of preferred spatial directions on cosmological scales. Thus,  
anomalies have been detected in the low multipoles of the CMB \cite{Ade:2015hxq,Schwarz:2015cma}. They include the alignment of quadrupole, octupole and ecliptic plane, a  dipole anomaly in the power spectrum that breaks statistical isotropy and the hemispherical anomaly whose maximum asymmetry is observed in the ecliptic frame. On the other hand, large scale bulk flows have also been detected with an amplitude which
has been claimed to exceed the predictions of standard $\Lambda$CDM \cite{Ade:2013opi,Atrio-Barandela:2014nda,Scrimgeour:2015khj}. Although the statistical significance of such anomalies is somewhat limited, they have motivated the search for preferred directions in cosmology. 

One of the simplest frameworks to explore
the consequences of  Lorentz symmetry breaking is the presence 
of tensor fields acquiring non-vanishing vacuum expectation values. This is indeed the case of the so called Standard Model Extension
(SME) \cite{Colladay:1998fq}. In particular, in the case in which such 
vacuum expectation value is acquired by a vector field, the first models were proposed by Nambu already in the sixties \cite{Nambu:1968qk}. Depending on the 
particular type of vector, this mechanism can induce two kinds of gravitational effects. On one hand, if the vector field is timelike, preferred frame effects would be present. On the other hand, a space-like VEV for the vector field 
will generate preferred directions effects in which we are interested in this paper.

Preferred frame effects have been explored in local gravitational experiments through the so called PPN formalism \cite{Will,Bailey:2006fd}. In particular, two PPN parameters, $\alpha_1$ and
$\alpha_2$, have been restricted by Solar System and pulsar observations. Also,  modifications in the gravity wave dispersion relations have been studied in \cite{Blas:2014aca}. 
From a theoretical point of view, theories of gravity such as Horava gravity \cite{Horava:2009uw} or Einstein-aether \cite{Jacobson:2000xp} have been shown to 
generate this kind of preferred frame effects. Also 
on the cosmological framework, different kinds of vector-tensor theories including temporal background vector 
fields have been analysed in the context of dark energy \cite{ArmendarizPicon,Boehmer,Beltran1,Beltran2,Beltran3,BGP,BGP2}.

Preferred directions effects have been explored in the framework
of the anisotropic PPN formalism \cite{K.Nordtvedt:1976zz} and bounds from 
laboratory experiments have been obtained in \cite{Muller:2007es}.
The possible cosmological implications have been studied both on the CMB temperature power spectrum \cite{Ade:2015hxq,Ackerman:2007nb} and in the matter distribution in \cite{Pullen,Jeong,Shiraishi,Tansella:2018hdm}.
In those works, the evolution both of the background and
perturbations is assumed to be the standard in $\Lambda$CDM  and the anisotropy is assumed to be present
only in the primordial power spectra. Such anisotropic
power spectrum  can be generated for instance in models of inflation with vectors \cite{Bartolo,Bartolo2,Joda3} or higher-spin fields \cite{Bartolo3}.
A different kind of effects would be those associated to the presence of non-comoving fluids singling  out a preferred direction as those considered in \cite{movingDE,quadrupole,hector}. 

However, in this work we will focus on a different possibility for the generation of preferred direction effects, i.e. that such directions are built in 
the theory of gravity itself. As commented before, theories of gravity involving additional vector degrees of freedom have been 
analyzed in detail in recent years in the case in which
the vector field acquires a temporal background.
If the background vector field is spatial, the
gravitational dynamics can give rise to a
modified  evolution of perturbations, thus introducing 
anisotropies in the corresponding transfer functions. Such modified evolution will however depend on the particular theory under consideration. 
It is precisely the aim of this 
work to analyze this kind of effects in a model-independent way
from the data that will be provided by future galaxy and weak-lensing 
surveys. 
With that purpose, we will consider the effective approach to modified gravity for theories involving vector degrees of freedom developed in \cite{Resco:2018ubr}. Within the sub-Hubble and quasi-static (QSA) approximations, which are  very well suited to 
galaxy surveys analysis, it is well-known that a very general modification of 
gravity 
involving additional scalar degrees of freedom can be described with only two additional parameters: an effective Newton constant $\mu(k,a) = G_{eff} / G$ and a gravitational slip parameter $\gamma(k,a)$ \cite{Pogosian:2010tj, Silvestri:2013ne}.
In the vector case, when the background vector field is purely temporal, the theory can still be parametrized only with $\mu(k,a)$ and $\gamma(k,a)$ parameters, but this scenario changes when we have a preferred direction.  In this case we need two additional  effective parameters (if dark matter vorticity
can be neglected as is usually the case) which relate matter density perturbations 
to vector and tensor metric perturbations.  Apart from the standard time $a$ and scale $k$ dependence,  those four effective 
parameters can have an
additional $x=\hat k\cdot \hat A$ dependence on the angle between the wave-vector direction $\hat{k}$ and the preferred direction fixed by the background vector field $\hat{A}$.

As mentioned above our goal is to analyze the impact of preferred directions effects in galaxy and weak lensing surveys. Future generations of galaxy maps such as J-PAS \cite{Benitez:2014ibt}, DESI \cite{Aghamousa:2016zmz} or Euclid \cite{Laureijs:2011gra}, will increase in a significant way the accuracy of cosmological parameter measurements. The two main observables that can be extracted from galaxy maps are, on one hand, the galaxy power spectrum \cite{Bernardeau:2001qr, Bassett:2009mm} and, on the other,  the weak lensing shear and convergence spectra \cite{Bartelmann:1999yn, Kilbinger:2014cea, Hu:1998az}.

The redshift-space galaxy power spectrum is the main observable for galaxy clustering \cite{Seo:2003pu}. It is sensitive to the growth of structures via the growth factor $ D (z) = \delta_m (z) / \delta_m (0) $. In addition, thanks to the Alcock-Paczynski effect \cite{Alcock:1979mp}, the power spectrum is sensitive to the Hubble parameter $ H (z) $ and the angular distance $ D_A (z) $. Finally, due to the peculiar velocities, the position of galaxies in redshift space are distorted (RSD) \cite{Samushia:2011cs}. This effect introduces a dependence on the line of sight that involves the growth function $ f (z) = d \ln D / d \ln (a) $. For all these reasons, the redshift space power spectrum has a strong dependence on the cosmological model and on the 
underlying gravitational theory. As a matter of fact, when a 
preferred direction is present, an 
additional $x$ dependence is present which can be disentangled from the standard
angular dependence induced by the RSD. Also the anisotropic effects generated
by the gravity modification could be distinguished from those
induced by anisotropic primordial power spectra. 

On the other hand, we have the weak lensing effect \cite{Bacon:2000sy, Kaiser:2000if} which is the distortion of the shape of galaxies due to the gravitational perturbations. For scalar perturbations, the possible distortions are the convergence $\kappa$, i.e  the change in the size of the image, and the shear $\gamma_1$ and $\gamma_2$, which modifies the ellipticity of the image. In the standard case, the shear power spectra can be obtained from the convergence power spectrum \cite{Kaiser:1996tp, Hu:1999ek}. Moreover, we have the following relationship between them,
\begin{align}\label{I.1}
P_{\gamma_1} + P_{\gamma_2} = P_{\kappa},
\end{align}
where, in principle, convergence and shear can be measured independently \cite{Kaiser:1992ps, Kaiser:1994jb}. These power spectra give us information about the gravitational perturbations that affect light propagation. When a preferred direction is present, density perturbations can source vector and tensor modes 
thus affecting the lensing distorsion tensor. In this case, a new effect is present which is the rotation $\omega$ of the images. This rotation mode is rarely studied in the literature because it is a higher-order effect in the standard $\Lambda$CDM cosmology \cite{Cooray:2002mj}. Also, to measure this rotation effect using  weak lensing surveys is not possible  because there is no information about the original orientation of the galaxy image \cite{Thomas:2016xhb}. However, as we will
show,  the rotation effect can be detected in an indirect way using the new closing relation,
\begin{align}\label{I.2}
P_{\gamma_1} + P_{\gamma_2} = P_{\kappa} + P_{\omega},
\end{align}
i.e. independent measurements of  $P_{\kappa}$, $P_{\gamma_1}$ and $P_{\gamma_2}$ will allow to  constrain the rotation power spectrum $P_{\omega}$. 
Moreover, the new $P_\omega$ cannot be generated by an
anisotropic primordial curvature spectrum, so that 
a violation of the closing relation (\ref{I.1})  
will be a smoking gun for this kind of
modifications of gravity.

Besides, we find that the modified convergence power spectrum acquires a line-of-sight dependence
which is absent in standard $\Lambda$CDM. This has allowed us  to construct the 
convergence multipole power spectrum. Thus a future detection
of a non-vanishing multipolar component could be a potential signal of the 
existence of a gravitational preferred direction.

The paper is organized as follows: in \ref{sec2} we briefly summarize the results of \cite{Resco:2018ubr} for the anisotropic modified gravity parametrization. In \ref{sec3} we analyze the multipole power spectrum of clustering in the presence of an anisotropic vector  background and we also study the effects of the anisotropy in the weak lensing signals. In \ref{sec4} we obtain the null geodesics in the presence of scalar, vector and tensor perturbations. In \ref{sec5} we calculate the distortion tensor, and in \ref{sec6} we compute the weak lensing power spectra using the model-independent parametrization. In \ref{sec7} we present the Fisher matrix analysis for the multipole power spectra case and we obtain the sensitivity for measurements of the effective $\mu(a,k,x)$ parameter. In \ref{sec8} we compute the Fisher matrix for the redshift space power spectrum of galaxies to compare with the multipole case. In \ref{sec9} we present the Fisher matrix of the convergence power spectrum and we obtain the sensitivity for the modified gravity parameters. In \ref{sec10} we apply the Fisher formalism to the case of an anisotropic primordial
curvature spectrum. In section \ref{sec11} we briefly discuss the results and conclusions. Finally in the Appendices we calculate the covariance matrices for the galaxy and convergence power spectra in the presence of preferred directions.

%%%%%%%%%%%%%%%%%%%%%%%%%%%%%%%%%%%%%%%%%%%%%%%%%%%%%%%%%%%%%%%%%%%%%%%%%%%%%%%%%%%%
\section{Model-independent parametrization of anisotropic modified gravities}\label{sec2}
%%%%%%%%%%%%%%%%%%%%%%%%%%%%%%%%%%%%%%%%%%%%%%%%%%%%%%%%%%%%%%%%%%%%%%%%%%%%%%%%%%%%

In this first section we summarize the results of \cite{Resco:2018ubr}
on the model-independent parametrization of modified gravity theories with an additional vector field $A_\mu$. Let us thus start by considering a general anisotropic Bianchi I cosmology with scalar $(\Phi,\Psi)$, vector $Q_i$ and tensor $h_{ij}$ perturbations in the longitudinal 
gauge \cite{Pereira},
\begin{align}\label{s3.1}
ds^{2}=a^{2} \, &\left[ -(1+2\Psi) \, d \tau^{2}+[(1-2\Phi) \, \Xi_{i j} + h_{i j}] \, dx^{i} dx^{j} \right. \nonumber\\
& \left. - 2 \, Q_{i} \, d \tau \, dx^{i} \right],
\end{align}
where $\Xi_{i j}$ is the Bianchi tensor that reduces to $\Xi_{i j} = \delta_{i j}$ in the isotropic limit and vector perturbation satisfy $k^i Q_i = 0$, whereas for 
tensors we have $k^i h_{i j} = 0$ and $h^i_i = 0$. Considering that the extra vector field $A_\mu$ can have both temporal and spatial background components, the equations that relate the
different metric and matter perturbations in the sub-Hubble regime and in the quasi-static approximation read in Fourier space,
\begin{align}\label{s1.39}
k^2 \, {\Psi} \equiv - 4 \pi G \, a^2 \, \rho \, \mu_{\Psi} \, \delta(k), 
\end{align}
\begin{align}\label{s1.40}
k^2 \, {\Phi} \equiv - 4 \pi G \, a^2 \, \rho \, \mu_{\Phi} \, \delta(k), 
\end{align}
\begin{align}\label{s1.41}
k^2 \, {Q}_i \equiv 16 \pi G \, a^2 \, \rho \, \mu_{Q} \,  \mathcal{A}_i \, \delta(k), 
\end{align}
\begin{align}\label{s1.42}
k^2 \, {h}_{i j} \equiv - 4 \pi G \, a^2 \, \rho \, \mu_{h} \, \Sigma_{i j} \, \delta(k), 
\end{align}
where we have neglected the contribution from dark matter vorticity.
Here $G$ is the gravitational Newton constant, $\rho$ is the pressureless matter density, $\delta(k)$ is the matter density contrast and,
\begin{align}\label{s1.43}
\mathcal{A}_i = \hat{A}_i-x \, \hat{k}_i, 
\end{align}
\begin{align}\label{s1.44}
\Sigma_{i j} = 2 \, \mathcal{A}_i \, \mathcal{A}_j-(1- x^2) \, (\delta_{i j}-\hat{k}_i \, \hat{k}_j), 
\end{align}
where hat denotes the corresponding unit vector and $x \equiv \hat k\cdot \hat A$. These quantities satisfy the following properties,
\begin{align}\label{s1.46}
&\Sigma_{i j} = \Sigma_{j i}, \,\,\, \hat{k}^{i} \Sigma_{i j} = 0, \nonumber\\ 
&\Sigma^i_{\,\,i} = 0, \,\,\, \hat{k}^i \mathcal{A}_i = 0,
\end{align}
Unlike the case of modified gravities with an additional scalar degree of freedom \cite{Silvestri:2013ne} four parameters $(\mu_{\Psi}, \mu_{\Phi}, \mu_{Q}, \mu_{h})$ are needed to describe the most general modification of gravity in the presence of vector and tensor perturbations and an anisotropic background.
Notice that in general such parameters are functions of $(a,k,x)$. In the particular case of an isotropic background $A_i = 0$, we need only two parameters $\mu_{\Psi}$ and $\mu_{\Phi}$ which are related to  
those of the scalar  case as $\mu = \mu_{\Psi}$ and $\gamma = \mu_{\Phi} / \mu_{\Psi}$ with

\begin{equation}\label{3a}
\mu = \frac{G_{eff}}{G},
\end{equation}

and 

\begin{equation}\label{3b}
\gamma = \frac{\Phi}{\Psi}.
\end{equation}
%

%%%%%%%%%%%%%%%%%%%%%%%%%%%%%%%%%%%%%%%%%%%%%%%%%%%%%%%%%%%%%%%%%%%%%%%%%%%%%%%%%%%%
\section{Galaxy power spectrum} \label{sec3}
%%%%%%%%%%%%%%%%%%%%%%%%%%%%%%%%%%%%%%%%%%%%%%%%%%%%%%%%%%%%%%%%%%%%%%%%%%%%%%%%%%%%

Now we will analyze how  the galaxy power spectrum is
modified for  the theories of gravity we have just introduced. We will consider for simplicity the case in which the background metric
is the standard Robertson-Walker metric of $\Lambda$CDM cosmology. In this case, 
the only effects of the preferred direction come either from the 
matter power spectrum which can now exhibit an statistical anisotropy 
$P=P(k,x)$ or from the growth of scalar perturbations. Indeed,  let us define the growth factor $D(a)$ normalized 
as $D(a)=\delta(a)/\delta(0)$ and the corresponding growth function
\begin{equation}\label{2}
f(z) = \frac{d\ln(D(a))}{d\ln(a)},
\end{equation}
being $a = 1/(1+z)$ the scale factor, which satisfies
\begin{equation}\label{5}
\dot{f}+f^{2}+\left(2+\frac{\dot{H}}{H}\right) \, f-\frac{3}{2} \, \mu \, \Omega_{m} (a)=0,
\end{equation}
where dots denotes derivative with respect to $\ln a$, $H(a) = H_{0} E(a)$ is the Hubble parameter and $\Omega_{m} (a)$ is the matter density parameter $\Omega_{m} (a)=\Omega_m \, a^{-3} \, \frac{H_{0}^{2}}{H^{2}(a)}$. Notice that the only modification
with respect to the standard cosmology is the appearance of the effective 
parameter $\mu(a,k,x)$ which introduces the scale and direction dependence in the
growth evolution. For small anisotropy we can always expand \cite{Resco:2018ubr},
\begin{equation}\label{4}
\mu(a,k,x) = \mu_0(a,k) + \mu_2(a,k) \, x^2 + \mu_4(a,k) \, x^4 + O(x^6).
\end{equation}

Taking the anisotropic growth into account, the redshift-space linear galaxy power spectrum can be written as \cite{Seo:2003pu},
\begin{eqnarray}\label{1}
P_g (z,k,\hat{\mu},x) &=& \left[ (1 + \beta(z,k,x) \, \hat{\mu}^2) \, b(z) \, D(z,k,x) \right]^2 \nonumber\\
 &\times& P(k,x),
\end{eqnarray}
where $k$ is the modulus of the perturbation $\vec{k}$, and $\hat{\mu} = \hat{k} \cdot \hat{n}$ being $\hat{n}$ the line of sight. Here $P(k,x)$ is the matter power spectrum today which can be related with the matter power spectrum today in $\Lambda$CDM model $P_\Lambda(k)$ as,
\begin{equation}\label{1a}
P(k,x) = \mathrm{exp}\left[\int_0^ {z_{mat}} \, \frac{f(z',k,x)-f_{\Lambda}(z')}{1+z'} dz'\right] \, P_{\Lambda} (k),
\end{equation}
where $f_{\Lambda}(z)$ is the growth function in $\Lambda$CDM and 
we have assumed that for $z>z_{mat}$, $f(z,k,x)=f_{\Lambda}(z)$. 
For the sake of concreteness in the forecast analysis we will assume that $z_{mat}=10$ although the results are not very sensitive to its precise value.
As mentioned before, $D(z,k,x)$ is the growth factor, $b(z)$ is the galaxy bias and $\beta(z,k,x) = f(z,k,x) / b(z)$. 

As we can see from (\ref{1}), the redshift-space galaxy power
spectrum has two different kinds of anisotropic contributions: on one hand
the standard contribution from redshift space distorsions
(RSD) which introduces a quadrupole and hexadecapole
in $\hat \mu$, and
on the other, an extra contribution coming from the $x$ 
dependence of the growth function. Thus performing a multipole expansion 
with  respect to the line of sight we find,
\begin{equation}\label{8}
P_g (z,k,\hat{\mu},x) = \sum_{\ell} \, P_{\ell} (z,k,x) \, \mathcal{L}_{\ell} (\hat{\mu})
,
\end{equation}
where $\mathcal{L}_{\ell}$ are the Legendre polynomials so that
\begin{equation}\label{9}
P_{\ell} (z,k,x) = \frac{2\ell+1}{2} \, \int_{-1}^{1} \, d\hat{\mu} \, P_g (z,k,\hat{\mu},x) \, \mathcal{L}_{\ell} (\hat{\mu}).
\end{equation}
obtaining $P_{\ell} (z,k,x)$ different from zero for $\ell = 0, 2, 4$ i.e. we recover the well-known monopole, quadrupole and hexadecapole contributions
but with the new $x$ dependence.

In the particular case in which the modified gravity parameter $\mu$  is time independent, i.e. $\mu=\mu(k,x)$  a simple analytical expression for $f(z, k,x)$ can be obtained \cite{Resco:2017jky},
\begin{equation}\label{6}
f(z,k,x) = \xi(\mu(k,x)) \, f_{\Lambda}(z),
\end{equation}
being $f_{\Lambda}(z) = \Omega_m^{\gamma} (z)$ with $\gamma = 0.55$ \cite{Linder:2005in, Linder:2007hg} and,
\begin{equation}\label{7}
\xi(\mu) = \frac{1}{4} \, (\sqrt{1 + 24 \,  \mu} - 1).
\end{equation}
In this case, explicit expressions for the multipoles can be obtained. Thus we have
\begin{eqnarray}\label{10}
P_{0} (z,k,x) &=& \left( 1 + \frac{2}{3} \, \xi(\mu) \, \beta_{\Lambda}(z) + \frac{1}{5} \, \xi^2(\mu) \, \beta_{\Lambda}^2(z) \right) \nonumber \\
&\times& b^2(z) \, D_{\Lambda}^{2 \, \xi(\mu)}(z) \, P(k,x),
\end{eqnarray}
\begin{eqnarray}\label{11}
P_{2} (z,k,x) &=& \left( \frac{4}{3} \, \xi(\mu) \, \beta_{\Lambda}(z) + \frac{4}{7} \, \xi^2(\mu) \, \beta_{\Lambda}^2(z) \right) \nonumber \\
&\times&
b^2(z) \, D_{\Lambda}^{2 \, \xi(\mu)}(z) \, P(k,x),
\end{eqnarray}
\begin{equation}\label{12}
P_{4} (z,k,x) = \frac{8}{35} \, \xi^2(\mu) \, \beta_{\Lambda}^2(z) \, b^2(z) \, D_{\Lambda}^{2 \, \xi(\mu)}(z) \, P(k,x),
\end{equation}
where, 
\begin{equation}\label{13}
\beta_{\Lambda} (z) = \frac{f_\Lambda(z)}{b(z)} \, ,
\end{equation}
and
\begin{equation}\label{14}
f_{\Lambda}(z) = \frac{d\log(D_{\Lambda}(a))}{d\log(a)}.
\end{equation}
The multipole coefficients $P_\ell$, depend 
 in turn on the angular variable $x$ and therefore could be  additionally expanded in a different multipole expansion
with respect to $x$. Alternatively, a bi-polar expansion in $(\hat{\mu},x)$ \cite{Shiraishi:2016wec}, could have been  performed. However for the Fisher analysis that we will perform in this work, we will directly work with the
$P_\ell$ coefficients.

%%%%%%%%%%%%%%%%%%%%%%%%%%%%%%%%%%%%%%%%%%%%%%%%%%%%%%%%%%%%%%%%%%%%%%%%%%%%%%%%%%%%
\section{Weak lensing: null geodesics with scalar, vector and tensor perturbations}\label{sec4}
%%%%%%%%%%%%%%%%%%%%%%%%%%%%%%%%%%%%%%%%%%%%%%%%%%%%%%%%%%%%%%%%%%%%%%%%%%%%%%%%%%%%

In order to obtain the convergence and shear power spectra for weak lensing in the 
presence of scalar, vector and tensor perturbations, we start with the Bianchi perturbed metric (\ref{s3.1}), where as in the previous section we have considered for simplicity $\Xi_{i j} \simeq \delta_{i j}$. We will also work in cosmological time $t$ so that the metric reads
\begin{align}\label{s1.1}
ds^{2}=& -(1+2\Psi) \, d t^{2}+a(t)^2 \, [(1-2\Phi) \, \delta_{i j} + h_{i j}] \, dx^{i} dx^{j} \nonumber\\
& - 2 \, Q_{i} \, a(t) \, d t \, dx^{i},
\end{align}
For this metric, we are interested in 
deriving the corresponding null geodesics, satisfying
\begin{align}\label{s1.2}
\frac{d^2 x^i}{d \lambda^2} + \Gamma_{\alpha \beta}^i \, \frac{d x^{\alpha}}{d \lambda} \, \frac{d x^{\beta}}{d \lambda} = 0.
\end{align}
We will consider the angular perturbation with respect to the line of sight induced by the metric perturbations.
Thus, we define $x^i = \chi \, \theta^i$ where $\chi=\chi(z)$ is the comoving radial distance and $\theta^i = (\theta^1, \theta^2, 1)$, so that $\theta^i$ for $i=1,2$ are first order in the gravitational perturbations and $x^3 = \chi$. The goal is to obtain the geodesics (\ref{s1.2}) for $i=1,2$,
\begin{align}\label{s1.3}
\frac{d^2 x^i}{d \lambda^2} = \frac{d \chi}{d \lambda} \, \frac{d}{d\chi} \, \left( \frac{d \chi}{d \lambda} \, \frac{d}{d\chi} \, \left(\chi \, \theta^i\right) \right) ,
\end{align}
where,
\begin{align}\label{s1.4}
\frac{d \chi}{d \lambda} = \frac{d \chi}{d t} \, \frac{d t}{d \lambda},
\end{align}
and $\frac{d \chi}{d t} = - \frac{1}{a}$. In order to obtain $\frac{d t}{d \lambda}$ we define $P^{\mu} = \frac{d x^{\mu}}{d \lambda}$, where, for null geodesics,
\begin{align}\label{s1.5}
g_{\mu \nu}P^{\mu}P^{\nu} = 0,
\end{align}
so, at order zero in perturbations, we have,
\begin{align}\label{s1.6}
-(P^0)^2 + g_{i j} P^i P^j = 0.
\end{align}
By defining $p^2 \equiv g_{i j} P^i P^j$ we find,
\begin{align}\label{s1.7}
\frac{d t}{d \lambda} = p,
\end{align}
so that we obtain $\frac{d \chi}{d \lambda} = - \frac{p}{a}$ and, since for $i=1,2$, $\theta^i$ is first order in 
perturbations, we can write
\begin{align}\label{s1.8}
\frac{d^2 x^i}{d \lambda^2} = -\frac{p}{a} \, \frac{d}{d\chi} \, \left( -\frac{p}{a} \, \frac{d}{d\chi} \, \left(\chi \, \theta^i\right) \right).
\end{align}
Thus we only need $p$ to zeroth order, which satisfies  $p \, a \propto const$ so that,
\begin{align}\label{s1.9}
\frac{d^2 x^i}{d \lambda^2} = p^2 \, \frac{d}{d\chi} \, \left( \frac{1}{a^2} \, \frac{d}{d\chi} \, \left(\chi \, \theta^i\right) \right) .
\end{align}
On the other hand, we have the Christoffel symbol term,
\begin{align}\label{s1.10}
\Gamma_{\alpha \beta}^i \, \frac{d x^{\alpha}}{d \lambda} \, \frac{d x^{\beta}}{d \lambda} = \left( \frac{d \chi}{d \lambda} \right)^2 \, \Gamma_{\alpha \beta}^i \, \frac{d x^{\alpha}}{d \chi} \, \frac{d x^{\beta}}{d \chi}.
\end{align}
For the metric (\ref{s1.1}), we have,
\begin{align}\label{s1.11}
\Gamma_{0 0}^i = a^{-2} \, \Psi_{,i} - a^{-1} \, \left( H \, Q_i + Q_{i , 0} \right),
\end{align}
\begin{align}\label{s1.12}
\Gamma_{j 0}^i = \delta_{i j} \, \left( H-\Phi_{,0} \right)-a^{-1} \, Q_{[i, j]} + \frac{1}{2} \, h_{i j , 0},
\end{align}
\begin{align}\label{s1.13}
\Gamma_{j k}^i = \,& \Phi_{,i} \, \delta_{j k} - \Phi_{,k} \, \delta_{i j} - \Phi_{,j} \, \delta_{k i} + a \, H \, Q_{i} \, \delta_{j k} \nonumber\\
& + \frac{1}{2} \, (h_{i j , k} + h_{i k , j} - h_{j k , i}),
\end{align}
where a comma denotes derivative with respect to the coordinates $(t, x^{1}, x^{2}, x^{3})$ and $H = \frac{1}{a} \, \frac{da}{dt}$ is the Hubble parameter. Let us analyze the different terms of equation (\ref{s1.10}):
\begin{itemize}
\item{$\alpha = \beta =0$ :} in this case we only have the term $\Gamma_{0 0}^i \, \left( \frac{d t}{d \chi} \right)^2$, and $\frac{d t}{d \chi} = -a$ to zeroth order, so that we obtain,
\begin{align}\label{s1.14}
\Gamma_{0 0}^i \, \left( \frac{d t}{d \chi} \right)^2 = \Psi_{, i} - a \, \left[ H \, Q_i + Q_{i , 0} \right].
\end{align}
\item{$\alpha = j, \, \beta =0$} (and the symmetric case): now we have $\Gamma_{j 0}^i \, \frac{d t}{d \chi} \, \frac{d x^j}{d \chi}$. For $j = 1,2$ the derivative $\frac{d x^j}{d \chi}$ is first order in perturbations, so that in this case $\Gamma_{j 0}^i$ must be order zero. However, when $j=3$, we have $\frac{d x^3}{d \chi} = 1$ then $\Gamma_{3 0}^i$ has to be first order in perturbations. Taking all the terms into account we obtain,
\begin{align}\label{s1.15}
\Gamma_{j 0}^i \, \frac{d t}{d \chi} \, \frac{d x^j}{d \chi} =& -a \, H \, \frac{d}{d \chi} (\chi \, \theta^i) \nonumber\\
&+ Q_{[i,3]} - \frac{1}{2} \, a \, h_{i3 , 0}.
\end{align}
 Notice that since we also have $\Gamma_{0 j}^i$, the term (\ref{s1.15}) contributes twice to the final expression.
\item{$\alpha = j, \, \beta =k$ :} finally we have $\Gamma_{j k}^i \, \frac{d x^j}{d \chi} \, \frac{d x^k}{d \chi}$, because $x^j$ is order one when $j \neq 3$ and $\Gamma_{j k}^i$ is always order one, the only term that contributes corresponds to $j=k=3$ ($i = 1,2$),
\begin{align}\label{s1.16}
\Gamma_{3 3}^i \, \left( \frac{d x^3}{d \chi} \right)^2 = \Phi_{,i} + a \, H \, Q_i + h_{i3,3} - \frac{1}{2} \, h_{33,i}.
\end{align}
\end{itemize}
As we can see in the previous analysis, $\Gamma_{\alpha \beta}^i \, \frac{d x^{\alpha}}{d \chi} \, \frac{d x^{\beta}}{d \chi}$ is first order in perturbations, so that the prefactor $\left( \frac{d \chi}{d \lambda} \right)^2$ in (\ref{s1.10}) must be of zeroth order. Finally, equation (\ref{s1.10}) becomes,
\begin{align}\label{s1.17}
\Gamma_{\alpha \beta}^i \, \frac{d x^{\alpha}}{d \lambda} \,& \frac{d x^{\beta}}{d \lambda} = \left( \frac{p}{a} \right)^2 \, \left[ (\Phi + \Psi)_{, i} - 2 \, a \, H \, \frac{d}{d \chi} (\chi \, \theta^i) \right. \nonumber\\
&\left. + 2 \, Q_{[i,3]} + h_{i3,3} - \frac{1}{2} \, h_{33 ,i} - a \, (Q_i + h_{i3})_{,0}   \right].
\end{align}
If we expand (\ref{s1.9}), and taking into account that $\frac{d}{d \chi} = - a^2 \, H \, \frac{d}{da}$, we find,
\begin{align}\label{s1.18}
\frac{d^2 x^i}{d \lambda^2} = \left( \frac{p}{a} \right)^2 \, \left[ \frac{d^2 (\chi \, \theta^i)}{d \chi^2} + 2 \, a \, H \, \frac{d}{d \chi} (\chi \, \theta^i) \right].
\end{align}
Thus, using (\ref{s1.17}) and (\ref{s1.18}) we can obtain from the geodesic equation (\ref{s1.2}),
\begin{align}\label{s1.19}
\frac{d^2}{d \chi^2}(\chi \, \theta^i) =& - (\Phi + \Psi)_{,i} - 2 Q_{[i,3]} - h_{i3,3} + \frac{1}{2} \, h_{33,i} \nonumber\\
&+a\,(Q_i + h_{i3})_{,0}.
\end{align}
At this point we apply the quasi-static approximation (QSA) and the sub-Hubble regime in which we can neglect the time derivatives of perturbations with respect to the spatial derivatives,
\begin{align}\label{s1.20}
\frac{d^2}{d \chi^2}(\chi \, \theta^i) = - (\Phi + \Psi)_{,i} - 2 Q_{[i,3]} - h_{i3,3} + \frac{1}{2} \, h_{33,i}.
\end{align}
It will be useful to define the source term of equation (\ref{s1.20}) as,
\begin{align}\label{s1.21}
Y_i \equiv - \left(\Phi + \Psi + Q_3 + \frac{1}{2} \, h_{33}\right)_{,i} - (Q_i + h_{i3})_{, 3},
\end{align}
In the following section we will proceed with the integration of equation (\ref{s1.20}) and the definition of the distortion tensor.

%%%%%%%%%%%%%%%%%%%%%%%%%%%%%%%%%%%%%%%%%%%%%%%%%%%%%%%%%%%%%%%%%%%%%%%%%%%%%%%%%%%%
\section{The distortion tensor}\label{sec5}
%%%%%%%%%%%%%%%%%%%%%%%%%%%%%%%%%%%%%%%%%%%%%%%%%%%%%%%%%%%%%%%%%%%%%%%%%%%%%%%%%%%%

By integrating twice equation (\ref{s1.20}) we obtain,
\begin{align}\label{s1.22}
\theta_{i}^{S} = \frac{1}{\chi} \, \int_0^{\chi} \, d \chi'' \, \int_0^{\chi''} \, d \chi' \, Y_i (\chi'\vec{\theta}) + \textrm{const}.
\end{align}
Since the integrand is just a function of $\chi'$, we can integrate over $\chi''$ and fix the integration constant as the initial angle $\theta_i$,
\begin{align}\label{s1.23}
\theta_{i}^{S} = \theta_i + \int_0^{\chi} \, d \chi' \, Y_i (\chi'\vec{\theta}) \, \left( 1-\frac{\chi'}{\chi} \right).
\end{align}
Now, we define the distortion tensor as,
\begin{align}\label{s1.24}
\psi_{i j} \equiv \frac{\partial \theta_i^{S}}{\partial \theta_j} - \delta_{i j}.
\end{align}
By using $\frac{\partial}{\partial \theta_j}=\frac{\partial x_k}{\partial \theta_j} \, \frac{\partial}{\partial x_k} = \chi \, \frac{\partial}{\partial x_j}$ we obtain,
\begin{align}\label{s1.25}
\psi_{i j} = \int_0^{\chi} \, d \chi' \, \chi' \, Y_{i, j} \, \left( 1-\frac{\chi'}{\chi} \right),
\end{align}
where $\psi_{i j} = \psi_{i j}(\chi, \vec{\theta})$. We want to integrate over $\chi$ to project onto the two-dimensional $(\theta_1, \theta_2)$ plane. In general the survey contains a distribution of galaxies $W(\chi)$, which is normalized as $\int_0^{\chi_\infty} d\chi W(\chi) = 1$, where $\chi_\infty=\lim_{z\rightarrow \infty}\chi(z)$ so that the projected distortion tensor is,
\begin{align}\label{s1.26}
\psi_{i j} (\vec{\theta}) = \int_0^{\chi_{\infty}} \, d\chi \, W(\chi) \, \int_0^{\chi} \, d \chi' \, \chi' \, Y_{i, j} \, \left( 1-\frac{\chi'}{\chi} \right).
\end{align}
By changing the order of integration, we can obtain,
\begin{align}\label{s1.27}
\psi_{i j} (\vec{\theta}) = \int_0^{\chi_{\infty}} d\chi \,\, \chi \, g(\chi) \, Y_{i, j} (\chi, \vec{\theta}),
\end{align}
where we have defined,
\begin{align}\label{s1.28}
g(\chi) \equiv \int_{\chi}^{\chi_{\infty}} \, d\chi' \, \left( 1-\frac{\chi}{\chi'} \right) \, W(\chi').
\end{align}
As we have seen, $i = 1,2$ so that $\psi_{i j}$ is a $2\times 2$ matrix. This matrix is non-symmetric in general as we can see in (\ref{s1.27}),
\[
   \psi_{i j} \equiv
  \left( {\begin{array}{cc}
   - \kappa - \gamma_1 & -\gamma_2-\omega \\
   -\gamma_2+\omega & - \kappa + \gamma_1 \\
  \end{array} } \right)
\]
Thus, the convergence and shear parameters are,
\begin{align}\label{s1.29}
\kappa = - \frac{\psi_{11}+\psi_{22}}{2},
\end{align}
\begin{align}\label{s1.30}
\gamma_1 = - \frac{\psi_{11}-\psi_{22}}{2},
\end{align}
\begin{align}\label{s1.31}
\gamma_2 = - \frac{\psi_{12}+\psi_{21}}{2},
\end{align}
whereas the rotation parameter corresponds to
\begin{align}\label{s1.32}
\omega = - \frac{\psi_{12}-\psi_{21}}{2},
\end{align}
Now, we use equation (\ref{s1.21}) into (\ref{s1.27}) so that
\begin{align}\label{s1.33}
\psi_{i j} (\vec{\theta}) =& -\int_0^{\chi_{\infty}} d\chi \,\, \chi \, g(\chi) \, \left( \Phi + \Psi + Q_3 + \frac{1}{2} \, h_{33} \right)_{,ij} \nonumber\\
& -\int_0^{\chi_{\infty}} d\chi \,\, \chi \, g(\chi) \, \left( Q_i + h_{i3} \right)_{,3j},
\end{align}
As we can see from the previous equation, the vector and tensor perturbations generate the rotation effect in the distortion tensor \cite{Thomas:2016xhb}. 
Since $x^3 = \chi$, we can integrate by parts the second integral to obtain,
\begin{align}\label{s1.34}
\int_0^{\chi_{\infty}} d\chi \,\, \chi \, g(\chi) &\, \left( Q_i + h_{i3} \right)_{,3j} = \cancel{\left.\left[ \chi \, g(\chi) \, (Q_i + h_{i3})_{,j} \right]\right|_{0}^{\chi_{\infty}}} \nonumber\\
& - \int_0^{\chi_{\infty}} d\chi \,\, \left( g + \chi \frac{dg}{d\chi} \right) \, \left( Q_i + h_{i3} \right)_{,j},
\end{align}
so that the distortion tensor becomes,
\begin{widetext}
\begin{align}\label{s1.35}
\psi_{i j} (\vec{\theta}) = -\int_0^{\chi_{\infty}} d\chi \,\, \chi \, g(\chi) \, \left[ \left( \Phi + \Psi +Q_3 +\frac{1}{2} \, h_{33} \right)_{,ij} - \frac{1}{\chi} \, \left( 1+\frac{\chi}{g} \, \frac{dg}{d \chi} \right) \, (Q_i + h_{i3})_{,j} \right].
\end{align}
\end{widetext}
Now, we want to go to the Fourier space of $\vec{\theta}$ so that we define,
\begin{align}\label{s1.36}
\tilde{\psi}_{i j} (\vec{\ell}) = \int \, d^2 \theta \,\, \textrm{e}^{-i \, \vec{\ell} \cdot \vec{\theta}} \, \psi_{i j} (\vec{\theta}).
\end{align}
Taking into account that,
\begin{align}\label{s1.37}
\frac{\partial}{\partial x^i} = \frac{1}{\chi} \, \frac{\partial}{\partial \theta^i},
\end{align}
the Fourier transform of the distorsion matrix is,
\begin{widetext}
\begin{align}\label{s1.38}
\tilde{\psi}_{i j} (\vec{\ell}) = \int_0^{\chi_{\infty}} d\chi \,\, \frac{g(\chi)}{\chi} \, \left[ \ell_i \, \ell_j \, \left( \tilde{\Phi} + \tilde{\Psi} +\tilde{Q}_3 +\frac{1}{2} \, \tilde{h}_{33} \right) + i \, \ell_j \, \left( 1+\frac{\chi}{g} \, \frac{dg}{d \chi} \right) \, (\tilde{Q}_i + \tilde{h}_{i3}) \right],
\end{align}
\end{widetext}
The power spectrum of this distortion matrix is the weak-lensing observable. In the following section we will relate vector and tensor perturbations to the matter density perturbations using (\ref{s1.39})-(\ref{s1.42}), so that we can obtain the final weak-lensing power spectrum in terms of the matter power spectrum. 

%%%%%%%%%%%%%%%%%%%%%%%%%%%%%%%%%%%%%%%%%%%%%%%%%%%%%%%%%%%%%%%%%%%%%%%%%%%%%%%%%%%%
\section{Weak-lensing power spectra}\label{sec6}
%%%%%%%%%%%%%%%%%%%%%%%%%%%%%%%%%%%%%%%%%%%%%%%%%%%%%%%%%%%%%%%%%%%%%%%%%%%%%%%%%%%%

The power spectrum of the distortion tensor is defined in the following way,
\begin{align}\label{s1.48}
P_{ijlm}^{\psi} (\vec{\ell}) \equiv \frac{1}{(2 \pi)^2} \, \int \, d^2 \ell' \, \langle \tilde{\psi}_{ij} (\vec{\ell}) \, \tilde{\psi}_{ij}^* (\vec{\ell}')  \rangle.
\end{align}
Using expressions (\ref{s1.39}) - (\ref{s1.42}), we can obtain the power spectrum (\ref{s1.48}) as a function of the matter power spectrum,
\begin{align}\label{s1.49}
\langle \delta(a, \vec{k}) \, \delta^*(a, \vec{k}') \rangle = (2 \pi)^3 \, \delta^3 (\vec{k} - \vec{k}') \, P(a, \vec k).
\end{align}
With that purpose it is first necessary to relate the Fourier transforms in the $\vec{\theta}$ and $\vec{x}$ variables. Let us thus denote with a bar the Fourier transform in $\vec x$ at a given time i.e.
\begin{align}\label{s1.47}
\bar{f} (\vec{k}) \equiv \int \, d^3 x \,\, \textrm{e}^{-i \, \vec{k} \cdot \vec{x}} \, f (\vec{x}).
\end{align}
Thus, we can write
\begin{align}\label{s1.50}
\bar{f}(k_3,\vec\ell) = \int d \chi \int \chi^2 \, d^2 \theta \, \textrm{e}^{-i \, \vec{\ell} \cdot \vec{\theta}} \, \textrm{e}^{-i \, k_3 \, \chi} \, f(\vec{x}),
\end{align}
where we have used $\ell_i = \chi \, k_i$ for $i=1,2$ so that using the definition of the Fourier transform in $\vec \theta$ in (\ref{s1.36}) we obtain,
\begin{align}\label{s1.51}
\bar{f}(k_3,\vec\ell) = \int d \chi \, \chi^2 \, \textrm{e}^{-i \, k_3 \, \chi} \, \tilde{f}(\chi,\vec\ell).
\end{align}
By performing the inverse transform in $k_3$ we get,
\begin{align}\label{s1.52}
\tilde{f}(\chi,\vec\ell) = \frac{1}{2 \pi \chi^2} \int d k_3 \, \textrm{e}^{i \, k_3 \, \chi} \, \bar{f}(k_3,\vec\ell).
\end{align}
In order to obtain the power spectrum in (\ref{s1.48}), we rewrite  equation (\ref{s1.38}) for $\tilde{\psi}_{i j} (\vec{\ell})$ in the following compact way,
\begin{align}\label{s1.53}
\tilde{\psi} (\vec{\ell}) = \int_{0}^{\chi_{\infty}} \, d\chi \,\, \sum_{\alpha} \, C_{\alpha} (\chi, \vec\ell) \, \tilde{f}_{\alpha} (\chi,\vec\ell),
\end{align}
where we have omitted the indices, $f_{\alpha}$ are the different metric perturbations and $C_{\alpha}$ the corresponding coefficients. Using this expression we obtain,
\begin{widetext}
\begin{align}\label{s1.54}
P^{\psi} (\vec{\ell}) = \frac{1}{(2 \pi)^2} \, \int \, d^2 \ell' \, \int_{0}^{\chi_{\infty}} \, d\chi \, \int_{0}^{\chi_{\infty}} \, d\chi' \, \sum_{\alpha, \beta} \, C_{\alpha} (\chi, \vec\ell) \, C_{\beta}^{*} (\chi', \vec\ell') \, \langle \tilde{f}_{\alpha} (\chi,\vec\ell) \, \tilde{f}_{\beta}^{*} (\chi',\vec\ell') \rangle,
\end{align}
\end{widetext}
and using (\ref{s1.52}) in $\langle \tilde{f}_{\alpha} (\chi,\vec\ell) \, \tilde{f}_{\beta}^{*} (\chi',\vec\ell') \rangle$, we obtain
\begin{widetext}
\begin{align}\label{s1.55}
\langle \tilde{f}_{\alpha} (\chi,\vec\ell) \, \tilde{f}_{\beta}^{*} (\chi',\vec\ell') \rangle = \frac{1}{2 \pi \chi^2 } \, \frac{1}{2 \pi \chi'^2 } \, \int \, dk_3 \, \textrm{e}^{i k_3 \chi} \, \int \, dk'_3 \, \textrm{e}^{-i k'_3 \chi'} \, \langle \bar{f}_{\alpha} (k_3,\vec\ell) \, \bar{f}_{\beta}^{*} (k_3',\vec\ell') \rangle.
\end{align}
\end{widetext}
As we can see from (\ref{s1.39})-(\ref{s1.42}), metric perturbations $\bar{f}$ can be related to the 
density perturbations according to the following generic form,
\begin{align}\label{s1.56}
\bar{f}_{\alpha} (k_3, \vec\ell) = B_{\alpha}(\vec k) \, \delta(\vec k),
\end{align}
where $k_i=\ell_i/\chi$ for $i=1,2$, so that, formally we obtain,
\begin{align}\label{s1.57}
\langle \bar{f}_{\alpha} (k_3,\vec\ell) \, \bar{f}_{\beta}^{*} (k_3',\vec\ell') \rangle =  B_{\alpha}(\vec k) B_{\beta}^*(\vec k') \, \langle \delta (\vec k) \, \delta^* (\vec k') \rangle. \nonumber\\
\end{align}
Using equation (\ref{s1.49}) and considering,
\begin{align}\label{s1.58}
\delta^3 (\vec{k}-\vec{k}') = \delta^2 \left( \frac{\vec{\ell}}{\chi} - \frac{\vec{\ell}'}{\chi'} \right) \, \delta (k_3 - k_3'),
\end{align}
we obtain,
\begin{align}\label{s1.59}
\langle \tilde{f}_{\alpha} (\chi,\vec\ell) \,& \tilde{f}_{\beta}^{*} (\chi',\vec\ell') \rangle = \frac{2 \pi}{\chi^2 \, \chi'^2} \, \delta^2 \left( \frac{\vec{\ell}}{\chi} - \frac{\vec{\ell}'}{\chi'} \right) \nonumber\\
&\int dk_3 \, \textrm{e}^{i \, k_3 \, (\chi - \chi')} \, B_{\alpha}(\vec k) B_{\beta}^*(\vec k') \, P(\vec k).
\end{align}
For small distortion angles $\theta$, we can consider $k_3 \ll k_1, k_2$ so that,
$\vec k\simeq \vec\ell/\chi$  
and accordingly,
\begin{align}\label{s1.61}
\langle \tilde{f}_{\alpha} (\chi,\ell) \, \tilde{f}_{\beta}^{*} (\chi',\ell') &\rangle = \frac{(2 \pi)^2}{\chi^2 \, \chi'^2} \, \delta^2 \left( \frac{\vec{\ell}}{\chi} - \frac{\vec{\ell}'}{\chi'} \right) \, \delta(\chi-\chi') \nonumber\\
&B_{\alpha}\left(\frac{\vec\ell}{\chi}\right) B_{\beta}^*\left(\frac{\vec\ell'}{\chi'}\right)\, P\left(\frac{\vec\ell}{\chi}\right).
\end{align}
Using this expression in (\ref{s1.54}) and writing $\delta^2 \left( \frac{1}{\chi} (\vec{\ell}-\vec{\ell}') \right)=\chi^2 \, \delta^2 (\vec{\ell}-\vec{\ell}')$,
we obtain
\begin{align}\label{s1.62}
P^{\psi} (\vec{\ell}) &= \int_0^{\chi_{\infty}} \, \frac{1}{\chi^2} \, P\left( \frac{\vec\ell}{\chi} \right) \nonumber\\
&\left[ \sum_{\alpha, \beta} \, C_{\alpha} (\chi, \vec\ell) C_{\beta}^* (\chi, \vec\ell) B_{\alpha} \left(\frac{\vec\ell}{\chi}\right) B_{\beta}^* \left(\frac{\vec\ell}{\chi}\right) \right] \, d \chi.
\end{align}
Finally changing from $\chi$ to the redshift variable $z = 1/(1+a)$, including explicitly the time dependence of the matter
power spectrum through the growth factor $D^2(z)$ and using $\left[4 \pi G a^2 \rho\right]^2 = \frac{9 H_0^4}{4} \, \Omega_m^2 \, (1+z)^2$  we get,
\begin{widetext}
\begin{align}\label{s1.63}
P_{i j l m}^{\psi} (\vec{\ell}) = \frac{9 \, H_0^4 \, \Omega_m^2}{4} \, \int_0^{\infty} d z \, \frac{(1+z)^2}{H(z)} \, g^2(z) \, \frac{\kappa_i \, \ell_j \, \kappa_l^* \, \ell_m}{\ell^4} \, D^2(z) \, P\left(\frac{\vec\ell}{\chi (z)}\right),
\end{align}
\end{widetext}
where
\begin{align}\label{s1.64}
\kappa_i \equiv \ell_i \, \alpha - i \, \left( 1+\frac{\chi \, g'}{\chi' \, g} \right) \, \textrm{v}_i,
\end{align}
\begin{align}\label{s1.65}
\alpha \equiv \mu_{\Psi} (1+\gamma) - 4 \mu_{Q} \mathcal{A}_3 + \frac{1}{2} \, \mu_{h} \, \Sigma_{3 3},
\end{align}
\begin{align}\label{s1.66}
\textrm{v}_i \equiv 4 \mu_{Q} \mathcal{A}_i - \mu_{h} \Sigma_{i 3}, 
\end{align}
being $\gamma \equiv \frac{\mu_{\Phi}}{\mu_{\Psi}}$, a prime denotes derivative with respect to redshift and 
\begin{align}\label{s1.67}
g(z) = \int_z^{\infty} \left( 1-\frac{\chi(z)}{\chi(z')} \right) \, n(z') \, dz', 
\end{align}
with $n(z)dz=W(\chi)d\chi$ and $n(z)$ the galaxy density function as a function of  redshift.

Now we can use expressions (\ref{s1.29})-(\ref{s1.32}) to construct the power spectra for convergence, shear and rotation,
\begin{align}\label{s1.68}
P_{\kappa} = \frac{1}{4} \, \left( P^{\psi}_{1111} + P^{\psi}_{2222} + P^{\psi}_{1122} + P^{\psi}_{2211} \right),
\end{align}
\begin{align}\label{s1.69}
P_{\gamma_1} = \frac{1}{4} \, \left( P^{\psi}_{1111} + P^{\psi}_{2222} - P^{\psi}_{1122} - P^{\psi}_{2211} \right),
\end{align}
\begin{align}\label{s1.70}
P_{\gamma_2} = \frac{1}{4} \, \left( P^{\psi}_{1212} + P^{\psi}_{2121} + P^{\psi}_{1221} + P^{\psi}_{2112} \right),
\end{align}
\begin{align}\label{s1.71}
P_{\omega} = \frac{1}{4} \, \left( P^{\psi}_{1212} + P^{\psi}_{2121} - P^{\psi}_{1221} - P^{\psi}_{2112} \right),
\end{align}
These expressions can be written in a more compact fashion by introducing the following variables. We define $\ell_1 \equiv \ell \, \Upsilon$ and $\ell_2 \equiv \ell \, \sqrt{1-\Upsilon^2}$ where,
\begin{align}\label{s1.72}
\Upsilon \equiv \frac{\hat{A}^i \ell_i}{\ell \, \sqrt{1-\hat{A}_3^2}}.
\end{align}
Considering the small-angle approximation $k_3 \ll k_1, k_2$, the conditions $\hat{k}^i Q_i = 0$ and $\hat{k}^i h_{i j} = 0$ imply,
\begin{align}\label{s1.73}
\ell^i \textrm{v}_i = 0.
\end{align}
Using this expression we can write $\textrm{v}_2$ as a function of $\textrm{v}_1$ and then we relate it with $\textrm{v}^2 \equiv \textrm{v}_1^2 + \textrm{v}_2^2$,
\begin{align}\label{s1.74}
\textrm{v}_1^2 = (1-\Upsilon^2) \, \textrm{v}^2.
\end{align}
Finally using (\ref{s1.72}) and (\ref{s1.73}) in the expressions of the power spectra (\ref{s1.68}) - (\ref{s1.71}) we obtain,
\begin{align}\label{s1.75}
P_{\kappa} = P_{\alpha},
\end{align}
\begin{align}\label{s1.76}
P_{\gamma_1} = (1-2 \, \Upsilon^2)^2 \, P_{\alpha} + 4 \, \Upsilon^2 \, (1-\Upsilon^2) \, P_{\textrm{v}},
\end{align}
\begin{align}\label{s1.77}
P_{\gamma_2} = 4 \, \Upsilon^2 \, (1-\Upsilon^2) \, P_{\alpha} + (1-2 \, \Upsilon^2)^2 \, P_{\textrm{v}},
\end{align}
\begin{align}\label{s1.78}
P_{\omega} = P_{\textrm{v}},
\end{align}
where $P_{\alpha}$ and $P_{\textrm{v}}$ are,
\begin{widetext}
\begin{align}\label{s1.79}
P_{\alpha} = \frac{9 \, H_0^4 \, \Omega_m^2}{4} \, \int_0^{\infty} d z \, \frac{(1+z)^2}{H(z)} \, g^2(z) \, \frac{\alpha^2}{4} \, D^2(z) \, P\left(\frac{\ell}{\chi (z)}\right),
\end{align}
\begin{align}\label{s1.80}
P_{\textrm{v}} = \frac{9 \, H_0^4 \, \Omega_m^2}{4} \, \int_0^{\infty} d z \, \frac{(1+z)^2}{H(z)} \, g^2(z) \, \left( 1+\frac{\chi \, g'}{\chi' \, g} \right)^2 \, \frac{\textrm{v}^2}{4 \, \ell^2} \, D^2(z) \, P\left(\frac{\ell}{\chi (z)}\right).
\end{align}
\end{widetext}
As we can see from equations (\ref{s1.75})-(\ref{s1.78}), we have the following closing relation,
\begin{align}\label{s1.103a}
P_{\gamma_1} + P_{\gamma_2} = P_{\kappa} + P_{\omega},
\end{align}
This is a useful relation since it allows to determine the
rotation power spectrum, which is not directly measurable
in lensing surveys, from shear and convergence measurements. 

We can use the expressions of $\mathcal{A}_i$ and $\Sigma_{i j}$ considering $\hat{k}_3 \ll 1$, and the definition of $\Upsilon$, to obtain expressions for $\alpha$ and $\textrm{v}^2$,
\begin{align}\label{s1.81}
\alpha = \mu_{\Psi} \, (1+\gamma) - 4 \, \xi \, \mu_{Q} + (2 \, \xi^2 +(1-\xi^2) \Upsilon^2 - 1) \, \frac{\mu_{h}}{2},
\end{align}
\begin{align}\label{s1.82}
\textrm{v}^2 = (4 \, \mu_{Q} -2 \, \xi \, \mu_{h})^2\, (1-\xi^2)(1-\Upsilon^2),
\end{align}
being $\xi \equiv \hat{A}_3$ with $-1 \leq \xi \leq 1$. Since $\hat{A}_3$ is the projection of $\hat{A}$ along the line of sight, we can perform a multipole expansion of 
$\alpha^2$ and $\textrm{v}^2$ above, using the Legendre polynomials in $\xi$. Thus for $\alpha^2=\sum_{r=0}^4 M_\alpha^r P_r(\xi)$ we have
\begin{align}\label{s1.83}
M_{\alpha}^{0} = \frac{1}{20} \, f_1^2 + \frac{1}{6} \, f_1 \, f_2 + \frac{1}{4} \, f_2^2 + \frac{16}{3} \, \mu_{Q}^2, 
\end{align}
\begin{align}\label{s1.84}
M_{\alpha}^{1} = -4 \, \left( \frac{3}{5} \, f_1 + f_2 \right) \, \mu_{Q},
\end{align}
\begin{align}\label{s1.85}
M_{\alpha}^{2} = \left( \frac{1}{7} \, f_1 + \frac{1}{3} \, f_2 \right) \, f_1 + \frac{32}{3} \, \mu_{Q}^2 
\end{align}
\begin{align}\label{s1.86}
M_{\alpha}^{3} = - \frac{8}{5} \, f_1 \, \mu_{Q},
\end{align}
\begin{align}\label{s1.87}
M_{\alpha}^{4} = \frac{2}{35} \, f_1^2
\end{align}
where $f_1 \equiv (2-\Upsilon^2) \, \mu_{h}$ and $f_2 \equiv 2 \, \mu_{\Psi} \, (1+\gamma) - (1-\Upsilon^2) \, \mu_{h}$.
On the other hand for $\textrm{v}^2=\sum_{r=0}^4 M_\textrm{v}^r P_r(\xi)$
\begin{align}\label{s1.88}
M_{\textrm{v}}^{0} = \frac{8}{15} \, \mu_{h}^2 + \frac{32}{3} \, \mu_{Q}^2,
\end{align}
\begin{align}\label{s1.89}
M_{\textrm{v}}^{1} = - M_{\textrm{v}}^{3} = - \frac{32}{5} \, \mu_{Q} \, \mu_{h},
\end{align}
\begin{align}\label{s1.90}
M_{\textrm{v}}^{2} = \frac{8}{21} \, \mu_{h}^2 - \frac{32}{3} \, \mu_{Q}^2,
\end{align}
\begin{align}\label{s1.92}
M_{\textrm{v}}^{4} = - \frac{32}{35} \, \mu_{h}^2.
\end{align}
With these definitions we obtain,
\begin{widetext}
\begin{align}\label{s1.93}
P_{\alpha}^{r} = \frac{9 \, H_0^4 \, \Omega_m^2}{4} \, \int_0^{\infty} d z \, \frac{(1+z)^2}{H(z)} \, g(z)^2 \, \frac{M_{\alpha}^r}{4} \, D(z)^2 \, P\left(\frac{\ell}{\chi (z)}\right), \; r=0,1,2,3,4
\end{align}
\begin{align}\label{s1.94}
P_{\textrm{v}}^{r} = \frac{9 \, H_0^4 \, \Omega_m^2}{4} \, \int_0^{\infty} d z \, \frac{(1+z)^2}{H(z)} \, g(z)^2 \, \left( 1+\frac{\chi \, g'}{\chi' \, g} \right)^2 \, \frac{(1-\Upsilon^2) \, M_{\textrm{v}}^{r}}{4 \, \ell^2} \, D(z)^2 \, P\left(\frac{\ell}{\chi (z)}\right), \; r=0,1,2,3,4
\end{align}
\end{widetext}
Finally, if we want to analyze the weak lensing signal at different redshift bins, we define the following window functions,
\begin{align}\label{s1.95}
g_i (z) = \int_z^{\infty} \, \left( 1-\frac{\chi(z)}{\chi(z')} \right) \, n_i (z') \, dz', 
\end{align}
where we consider  a galaxy density function of the form,
\begin{align}\label{s1.96}
n(z) = \frac{3}{2 \, z_p^3} \, z^2 \, e^{-(z/z_p)^{3/2}}, 
\end{align}
being $z_p = z_{mean}/\sqrt{2}$ and $z_{mean}$ the survey mean redshift. Then, for each bin we have the following galaxy distribution function, where we have take into account the photometric redshift error $\sigma_i$ in the 
corresponding bin,
\begin{align}\label{s1.97}
n_i (z) \propto \int_{\bar{z}_{i-1}}^{\bar{z}_{i}} \, n(z') \, e^{\frac{(z'-z)^2}{2 \, \sigma_i^2}} \, dz', 
\end{align}
where  $\sigma_i = \delta z \, (1+z_i)$, $\bar{z}_i$ is the upper limit of the $i$-bin and $n_i (z)$ is normalized  to one. 

With these definitions, the convergence, shear and rotation multipole power spectra are,
\begin{widetext}
\begin{align}\label{s1.99}
P_{\kappa \,\, i j}^{\,r} (\ell, \Upsilon) = \frac{9 \, H_0^4 \, \Omega_m^2}{4} \, \int_0^{\infty} d z \, \frac{(1+z)^2}{H(z)} \, g_i(z) \, g_j(z) \, \frac{M_{\alpha}^r}{4} \, D(z)^2 \, P\left(\frac{\ell}{\chi (z)}\right),
\end{align}
\begin{align}\label{s1.100}
P_{\omega \,\, i j}^{\,r} (\ell, \Upsilon) = \frac{9 \, H_0^4 \, \Omega_m^2}{4} \, \int_0^{\infty} d z \, \frac{(1+z)^2}{H(z)} \, g_i(z) \, g_j(z) \, \left( 1+\frac{\chi \, g_i'}{\chi' \, g_i} \right) \, \left( 1+\frac{\chi \, g_j'}{\chi' \, g_j} \right) \, \frac{(1-\Upsilon^2) \, M_{\textrm{v}}^{r}}{4 \, \ell^2} \, D(z)^2 \, P\left(\frac{\ell}{\chi (z)}\right),
\end{align}
\begin{align}\label{s1.101}
P_{\gamma_1 \,\, i j}^{\,r} (\ell, \Upsilon) = (1-2 \, \Upsilon^2)^2 \, P_{\kappa \,\, i j}^{r} (\ell, \Upsilon) + 4 \, \Upsilon^2 \, (1-\Upsilon^2) \, P_{\omega \,\, i j}^{r} (\ell, \Upsilon),
\end{align}
\begin{align}\label{s1.102}
P_{\gamma_2 \,\, i j}^{\,r} (\ell, \Upsilon) = 4 \, \Upsilon^2 \, (1-\Upsilon^2) \, P_{\kappa \,\, i j}^{r} (\ell, \Upsilon) + (1-2 \, \Upsilon^2)^2 \, P_{\omega \,\, i j}^{r} (\ell, \Upsilon).
\end{align}
\end{widetext}

%%%%%%%%%%%%%%%%%%%%%%%%%%%%%%%%%%%%%%%%%%%%%%%%%%%%%%%%%%%%%%%%%%%%%%%%%%%%%%%%%%%%
\section{Fisher analysis for the multipole power spectrum}\label{sec7}
%%%%%%%%%%%%%%%%%%%%%%%%%%%%%%%%%%%%%%%%%%%%%%%%%%%%%%%%%%%%%%%%%%%%%%%%%%%%%%%%%%%%

Considering a set of cosmological parameters $\{p_\alpha\}$, the Fisher matrix for the multipole power spectrum (\ref{9}) can be written as \cite{Taruya:2011tz},
\begin{equation}\label{15}
F_{\alpha \beta} = \sum_{n, n'} \sum_{\ell, \ell'} \left.\frac{\partial P_{\ell}(\vec{k}_n)}{\partial p_{\alpha}}\right|_{f} \, C_{\ell \ell'}^{-1} (\vec{k}_n, \vec{k}_{n'}) \, \left.\frac{\partial P_{\ell'}(\vec{k}_{n'})}{\partial p_{\beta}}\right|_{f},
\end{equation}
where sub-index $f$ denotes that the corresponding quantity is evaluated on the fiducial model, $\vec{k}_n$ are the discrete modes and $C_{\ell \ell'} (\vec{k}_n, \vec{k}_{n'})$ is the covariance matrix. In Appendix A the explicit calculation of the covariance matrix for the anisotropic power spectrum can be found.  In each 
redshift bin  this expression reads,
\begin{widetext}
\begin{equation}\label{19}
F_{\alpha \beta} (z) = \frac{V(z)}{8 \pi^2} \, \int^{k_{max}}_{k_{min}} \, \int^{1}_{-1} \, k^2 \, dk \, dx \, \sum_{\ell,\ell'} \, \left.\frac{\partial P_{\ell}(k,z,x)}{\partial p_{\alpha}}\right|_{f} \, C_{\ell \ell'}^{-1} (z,k) \, \left.\frac{\partial P_{\ell'}(k,z,x)}{\partial p_{\beta}}\right|_{f},
\end{equation}
\begin{equation}\label{20}
C_{\ell \ell'} (z,k) = \frac{(2\ell+1)(2\ell'+1)}{2} \, \int_{-1}^{1} \, d\hat{\mu} \, \mathcal{L}_{\ell} (\hat{\mu}) \, \mathcal{L}_{\ell'} (\hat{\mu}) \, \left[ \left.P_g (z,k,\hat{\mu})\right|_{f} \, e^{-k^{2} \, \hat{\mu}^{2} \, \sigma_{r}^{2}} + \frac{1}{n(z)} \right]^2,
\end{equation}
\end{widetext}
where we have included the effect of  redshift errors \cite{Seo:2003pu} in the power spectrum through the  $e^{-k^{2} \, \hat{\mu}^{2} \, \sigma_{r}^{2}}$
factor, where $\sigma_{r}=(\delta z \, (1+z))/H(z)$ with $\delta z$ the redshift error. Here $n(z)$ is the mean galaxy density and $V(z)$ the volume of the bin $z$. We consider a flat fiducial model,
\begin{equation}\label{21}
V(z)=\frac{4 \pi \, f_{sky}}{3} \, \left[\chi^{3}(z+\Delta z/2)-\chi^{3}(z-\Delta z/2)\right],
\end{equation}
being $f_{sky}$ the fraction of the sky, $\Delta z$ the width of the bin and $\chi(z)$ the comoving radial distance,
\begin{equation}\label{22}
\chi(z)=H_{0}^{-1} \, \int_{0}^{z} \frac{dz'}{E(z')}.
\end{equation}
As expected, this Fisher matrix reduces to the isotropic case when $P_{\ell}(k,z,x) = P_{\ell}(k,z)$. Finally, we need to know the values for $k_{\text{min}}$ and $k_{\text{max}}$ in each bin. $k_{\text{min}}$ can be fixed to 0.007 $h$/Mpc \cite{Amendola:2013qna},  and we obtain $k_{\text{max}}=k_{\text{max}}(z_{a})$ by imposing that $\sigma^2(z_{a},\pi/2k_{\text{max}}(z_{a}))=0.35$ so that we only consider modes in the linear regime. Thus, the amplitude of 
the fluctuations at a scale $R$ at redshift $z$ is
given by
\begin{eqnarray}\label{23}
\sigma^{2}(z,R)=D^2(z)\int{\frac{k'^{2}\, dk'}{2\pi^{2}} P(k') |\hat{W}(R,k')|^{2}} ,
\end{eqnarray}
where we have used a top-hat filter $\hat{W}(R,k)$, defined by
\begin{equation}\label{24}
\hat{W}(R,k)=\frac{3}{k^{3}R^{3}} \, [\sin(kR)-kR\cos(kR)].
\end{equation}
\begin{table}[htbp]
\begin{tabular}{|c|c|c|c|c|}
\hline
$z$ & $k_{max}$ & $n \, \times \, 10^{-3}$ & $\delta \mu_0 / \mu_0 (\%)$ & $\delta \mu_2 / \mu_0 (\%)$ \\
\hline \hline 
0.6  &  0.195  &  3.56  &  1.44  &  2.33  \\ \hline
0.8  &  0.225  &  2.42  &  1.00  &  1.59  \\ \hline
1.0  &  0.260  &  1.81  &  0.74  &  1.14  \\ \hline
1.2  &  0.299  &  1.44  &  0.70  &  1.07  \\ \hline
1.4  &  0.343  &  0.99  &  0.68  &  1.02  \\ \hline
1.8  &  0.447  &  0.33  &  0.71  &  1.06  \\ \hline
\end{tabular}
\caption{Redshift bins, $k_{max}$ values in  $\textrm{h/Mpc}$ units, galaxy densities in $(\textrm{h/Mpc})^{3}$ units and relative errors for $\mu_0$ and $\mu_2$ for an Euclid-like survey. We compare $\delta \mu_2$ with respect to $\mu_0$ because the fiducial value of $\mu_2$ is zero.}
\label{Table1}
\end{table}
\begin{table}[htbp]
\begin{tabular}{|c|c|c|c|c|c|}
\hline
$z$ & $k_{max}$ & $n \, \times \, 10^{-3}$ & $\delta \mu_0 / \mu_0 (\%)$ & $\delta \mu_2 / \mu_0 (\%)$ & $\delta \mu_4 / \mu_0 (\%)$ \\
\hline \hline
0.6  &  0.195  &  3.56  &  1.64  &  8.16  &  9.12  \\ \hline
0.8  &  0.225  &  2.42  &  1.14  &  5.55  &  6.21  \\ \hline
1.0  &  0.260  &  1.81  &  0.83  &  4.00  &  4.48  \\ \hline
1.2  &  0.299  &  1.44  &  0.79  &  3.73  &  4.17  \\ \hline
1.4  &  0.343  &  0.99  &  0.76  &  3.56  &  3.98  \\ \hline
1.8  &  0.447  &  0.33  &  0.79  &  3.69  &  4.13  \\ \hline
\end{tabular}
\caption{The same as in Table I but including  $\mu_4$ as an 
additional independent parameter in the Fisher analysis.}
\label{Table2}
\end{table}
\begin{figure} %[h!]
  	\includegraphics[width=0.4975\textwidth]{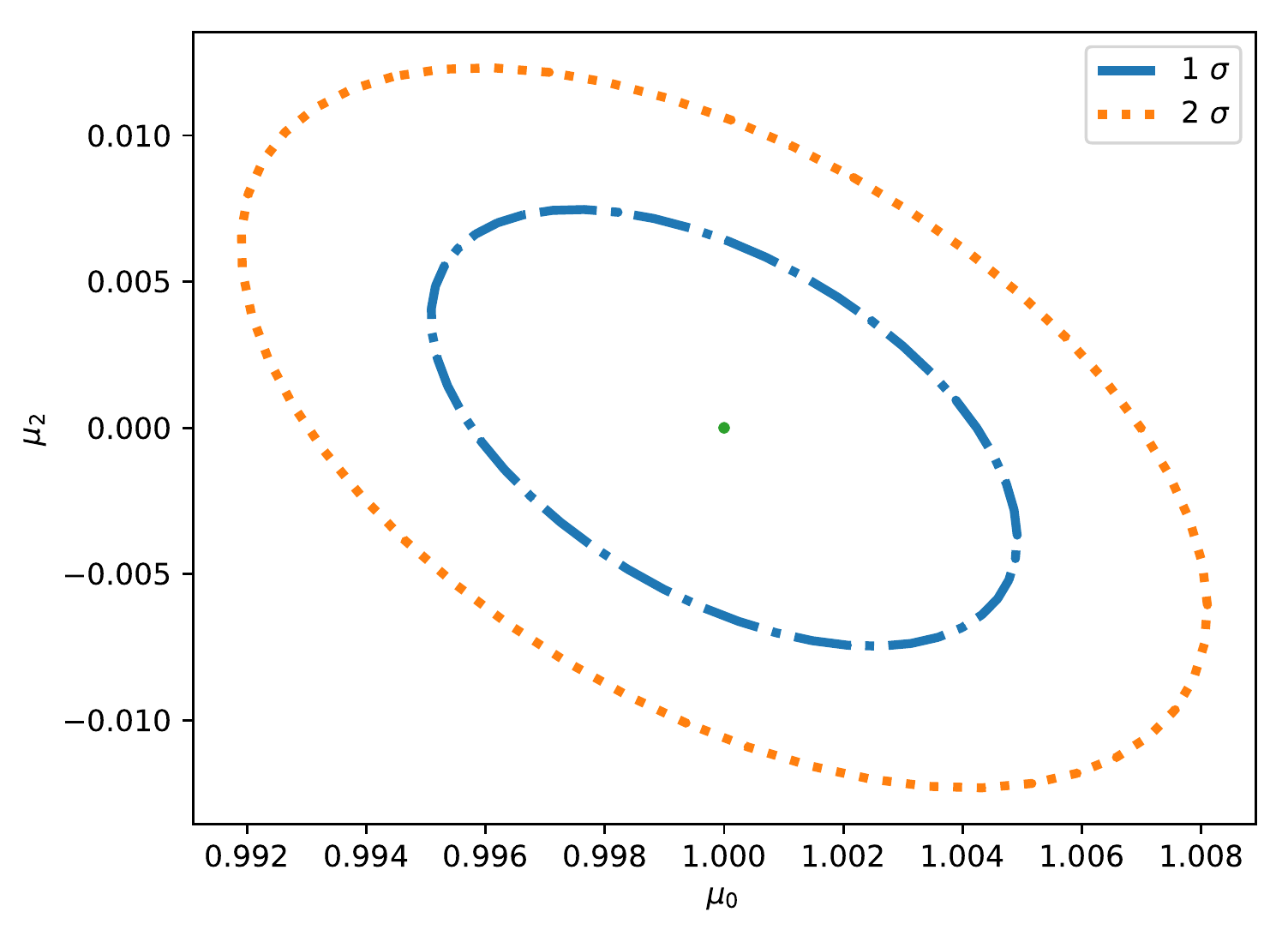}
		\caption{\footnotesize{Marginalized 1$\sigma$ and 2$\sigma$  regions for $\mu_0$ and $\mu_2$ for an Euclid-like survey from the multipole power spectrum information.}}
  \label{Figure_1}
\end{figure}
\begin{figure} %[h!]
  	\includegraphics[width=0.4975\textwidth]{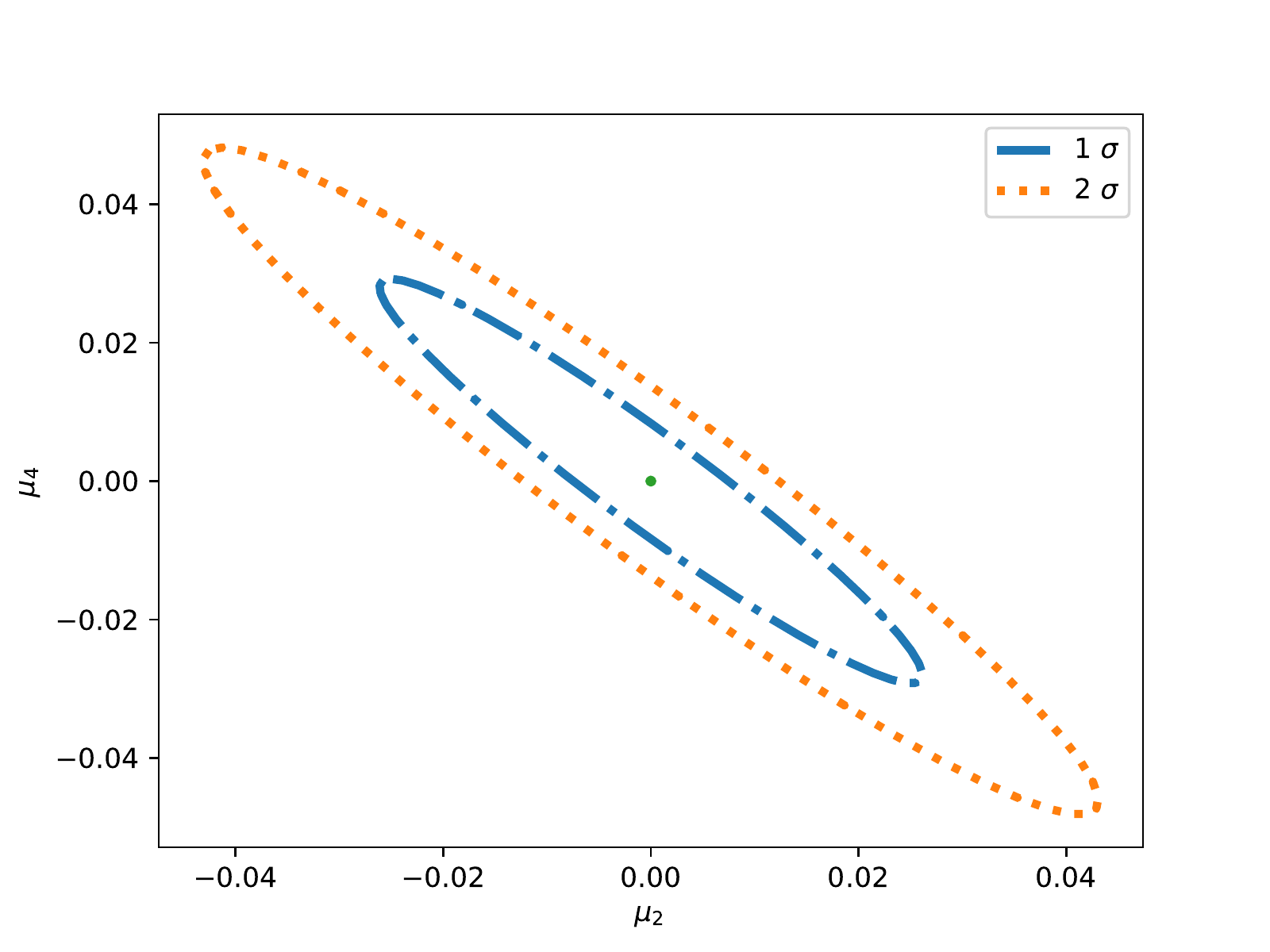}
		\caption{\footnotesize{Marginalized 1$\sigma$ and 2$\sigma$ regions for $\mu_2$ and $\mu_4$ for an Euclid-like survey from the multipole power spectrum information.}}
  \label{Figure_2}
\end{figure}
%

%%%%%%%%%%%%%%%%%%%%%%%%%%%%%%%%%%%
\subsubsection{Fiducial cosmology and galaxy redshift survey}
%%%%%%%%%%%%%%%%%%%%%%%%%%%%%%%%%%%

The fiducial cosmology we consider is given by $\Omega_{c} \, h^{2}=0.121$, $\Omega_{b} \, h^{2}=0.0226$, $\Omega_{\nu} \, h^{2}=0.00064$, $n_{s}=0.96$, $h=0.68$, $H^{-1}_{0}=2997.9 \, \textrm{Mpc/h}$, $\Omega_{k}=0$ and $\sigma_{8}=0.82$ in the standard \textrm{$\Lambda$CDM} model. For this cosmology,
\begin{equation}\label{26}
E(z)=\sqrt{\Omega_{m} \, (1+z)^{3}+(1-\Omega_{m})}.
\end{equation}
The growth function follows equation (\ref{6}) and the growth factor is,
\begin{equation}\label{27}
D_{\Lambda}(z)=\exp\left[\int_{0}^{N(z)} f_{\Lambda}(N') \, dN'\right],
\end{equation}
being $N(z)=-\ln(1+z)$. For the fiducial cosmology we obtain the present matter power spectrum $P(k)$ from CLASS \cite{Lesgourgues:2011re}. For the bias, we use a fiducial value of the form \cite{Laureijs:2011gra},
\begin{equation}\label{28}
b(z)=\sqrt{1+z}.
\end{equation}
Finally, we will limit ourselves to constant modified gravity parameters so that $\mu=\mu_0+\mu_2x^2+\mu_4x^4$, and since 
 we use \textrm{$\Lambda$CDM} as fiducial cosmology, we will take $[\mu_0, \mu_2, \mu_4]|_{r} = [1, 0, 0]$.

The inputs we need to compute $F_{\alpha \beta}$ are therefore redshift bins and the galaxy densities for each bin which can be found in Table \ref{Table1} and \ref{Table2} for an Euclid-like galaxy redshift survey. The fraction of the sky is $f_{sky}=0.364$ corresponding to 15000 $\mathrm{deg}^2$ and the redshift error is $\delta z=0.001$. 

First of all, we consider as independent parameters in each bin $[\beta_{\Lambda}, \mu_0, \mu_2]$ and we present the marginalized errors for $\mu_0$ and $\mu_2$ in Table \ref{Table1}. In Fig. \ref{Figure_1} we plot the 1-sigma and 2-sigma contours summing all the information in the whole redshift range. In such a case we obtain errors for $\mu_0$ and $\mu_2$ of order $1 \, \%$. 

Then, we add the parameter $\mu_4$ in each bin and we present the marginalized errors for $\mu_0$, $\mu_2$ and $\mu_4$ in Table \ref{Table2}. In Fig. \ref{Figure_2} we plot the 1-$\sigma$ and 2-$\sigma$ contours for $\mu_2$ and $\mu_4$ summing all the information in the full redshift range. As we can see, if we add a $x^4$ dependence,  the errors for $\mu_2$ increase in a factor $3-4$ but the errors for $\mu_0$ remain the same. Errors for $\mu_4$ are slightly larger than for $\mu_2$.
\\
%%%%%%%%%%%%%%%%%%%%%%%%%%%%%%%%%%%%%%%%%%%%%%%%%%%%%%%%%%%%%%%%%%%%%%%%%%%%%%%%%%%%
\section{Fisher analysis for the redshift-space power spectrum}\label{sec8}
%%%%%%%%%%%%%%%%%%%%%%%%%%%%%%%%%%%%%%%%%%%%%%%%%%%%%%%%%%%%%%%%%%%%%%%%%%%%%%%%%%%%

An alternative way to perform the Fisher analysis consists in using the redshift-space power spectrum (\ref{1}) rather than the multipoles considered in the previous section.
This, in fact,  allows  to take into account the Alcock-Paczynski effect \cite{Alcock:1979mp} so that we can write
\begin{equation}\label{29}
P_g (z,k_{r},\hat{\mu}_{r},\xi) = \frac{D_{A \, r}^{2} \, E}{D_{A}^{2} \, E_{r}} \, \left[ (1 + \beta_{\Lambda} \, \xi \, \hat{\mu}^2) \, b(z) \, D_{\Lambda}^{\xi} \right]^2 \, P(k),
\end{equation}
where as mentioned before, the $r$ sub-index denotes that the 
corresponding quantity is evaluated on the fiducial cosmology.
The dependence $k=k(k_{r})$ and $\hat{\mu}=\hat{\mu}(\hat{\mu}_{r})$ are given by,
\begin{equation}\label{30}
k=Q \, k_{r},
\end{equation}
\begin{equation}\label{31}
\hat{\mu}=\frac{E \, \hat{\mu}_{r}}{E_{r} \, Q},
\end{equation}
with
\begin{equation}\label{32}
Q=\frac{\sqrt{E^{2} \, \chi^{2} \, \hat{\mu}^{2}_{r}-E_{r}^{2} \, \chi_{r}^{2} \, (\hat{\mu}^{2}_{r}-1)}}{E_{r} \, \chi}.
\end{equation}
The $\xi = \xi(\mu)$ parameter follows equation (\ref{7}) and $\mu$ equation (\ref{4}). $D_{A}$ is the angular distance which, in a flat Universe, satisfies $D_A = (1+z)^{-1} \, \chi(z)$. 

\begin{figure*} %[h!]
 \begin{center}
 \includegraphics[width=0.45\textwidth]{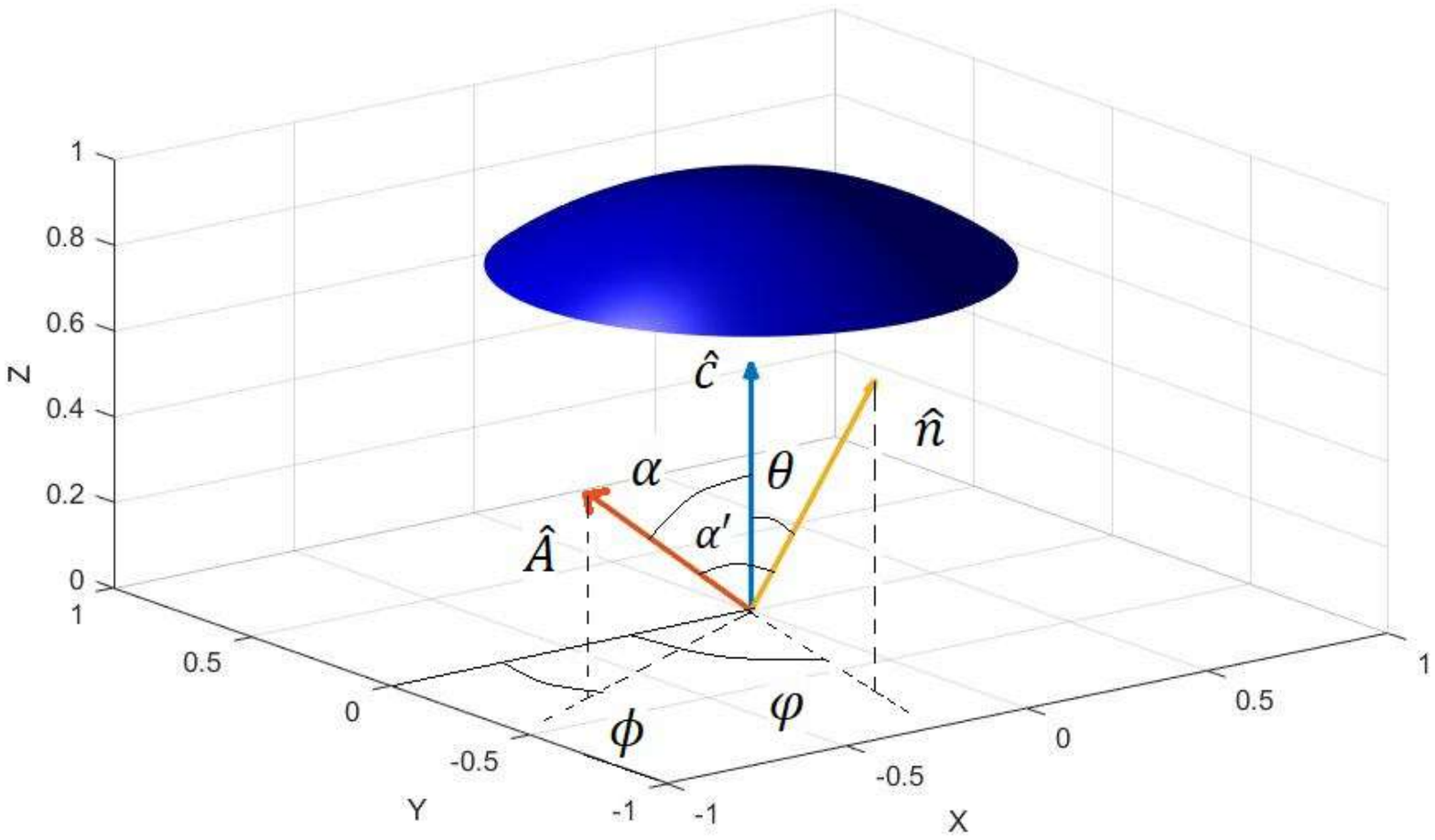}
 \includegraphics[width=0.45\textwidth]{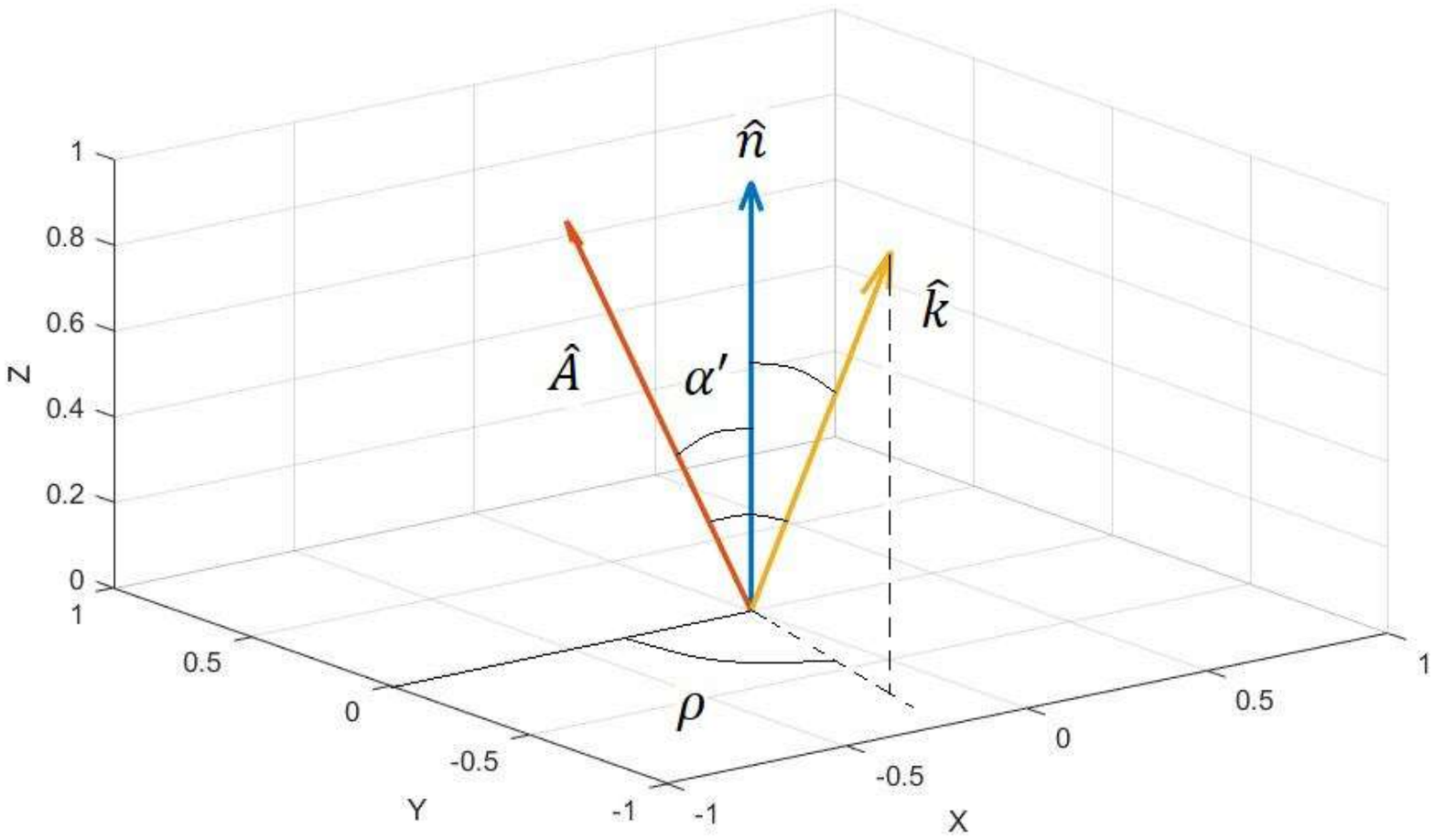} 
  \end{center}
  \caption{From left to right, reference frame for the Fisher analysis of the redshift space power spectrum, and auxiliary reference frame to calculate the integral in $\vec{k}$.}
 \label{Figure_3} 
 \end{figure*}

Thus, considering a set of cosmological parameters $\{p_\alpha\}$, the corresponding Fisher matrix for clustering at a given redshift bin centered at $z_a$ and for a solid angle of the survey centered at the line of sight $\hat{n}$ is,
\begin{eqnarray}\label{33}
dF_{i j}&= &\frac{1}{2} \, \int \, \frac{d^3 k}{(2 \pi)^3} \, \left.\frac{\partial \log P_g}{\partial p_{i}}\right|_{f} \, \left.\frac{\partial \log P_g}{\partial p_{j}}\right|_{f} \, \nonumber \\
&\times&\left. \left[ \frac{\bar{n} P_g}{1+\bar{n} P_g} \right] \right|_{f} \, dV_s,\nonumber
\end{eqnarray}
where,
\begin{equation}\label{34}
dV_s = V_z \, d\varphi \, d\theta \, \sin \theta,
\end{equation}
and
\begin{equation}\label{35}
V_z = \frac{1}{3} \, \left[\chi^{3}(z_a+\Delta z/2)-\chi^{3}(z_a-\Delta z/2)\right],
\end{equation}
with $\hat n(\theta,\varphi)$ where $\varphi$ and $\theta$ are the azimuthal and polar angles in the axes frame on the left panel of Fig. \ref{Figure_3}.

Since we are interested in summing all the angular information, we have to integrate over the angles $\varphi$ and $\theta$ but taking into account that $\left.\frac{\partial \log P_g}{\partial p_{\alpha}}\right|_{f}$ may depend on these angles. Thus, we integrate a spherical cap that encloses a fraction $f_{sky}$ of the sky,
\begin{widetext}
\begin{equation}\label{36}
F_{i j} = \frac{1}{2} \, \int \, \frac{d^3 k}{(2 \pi)^3} \, \int_0^{2\pi} \, d\varphi \, \int_{0}^{\arccos(1-2f_{sky})} \, \sin \theta \, d\theta \, \left.\frac{\partial \log P_g}{\partial p_{i}}\right|_{f} \, \left.\frac{\partial \log P_g}{\partial p_{j}}\right|_{f} \, \left. \left[ \frac{\bar{n} P_g}{1+\bar{n} P_g} \right] \right|_{f} \, V_z.
\end{equation}
\end{widetext}
The only angular dependences we have are $\hat{\mu} = \hat{k} \cdot \hat{n}$ and $x = \hat{k} \cdot \hat{A}$. It is useful to keep $\hat{\mu}$ as an integration variable, so that we have to relate $x$ with $\hat{\mu}$. With the choice of 
axes of Fig. \ref{Figure_3}, we find that,
\begin{equation}\label{37}
x = \sin \alpha' \, \sqrt{1-\hat{\mu}^2} \, \cos \rho + \cos \alpha' \, \hat{\mu},
\end{equation}
and,
\begin{equation}\label{38}
\cos \alpha' = \sin \alpha \, \sin \theta \, \cos (\varphi - \phi) + \cos \alpha \, \cos \theta,
\end{equation}
being  $\hat{A}(\alpha,\phi)$ with $\alpha$ and $\phi$ the polar and azimuthal
angles in the axes frame in Fig. 3 left, where the $Z$ axis is chosen in the direction of the center of the survey $\vec{c}$. 
 Thus, $x = x(\alpha, \phi, \hat{\mu}, \rho, \theta, \varphi)$ so that we have the following integration variables $[k, \hat{\mu}, \rho, \theta, \varphi]$. Finally, we have chosen as independent parameters for the Fisher matrix in each bin: $[E, D_{\Lambda}, \mu_0, \mu_2, \mu_4]$. For these parameters the derivatives are, 
\begin{eqnarray}\label{39}
\left.\frac{\partial \log P_g}{\partial E}\right|_{f}&=&1+\frac{2 \, \Delta z_{a}}{E^{2} \,H_{0} \, \chi(z_{a})}\nonumber\\
&+&\frac{4\beta_{\Lambda}\hat{\mu}^{2}(1-\hat{\mu}^{2})}{1+\beta_{\Lambda}\,\hat{\mu}^{2}} \left( \frac{1}{E}-\frac{\Delta z_{a}}{E^{2} \,H_{0} \, \chi(z_{a})} \right),\nonumber\\
\end{eqnarray}
\begin{eqnarray}\label{40}
\left.\frac{\partial \log P_g}{\partial D_{\Lambda}}\right|_{f} = \frac{2}{D_{\Lambda}},
\end{eqnarray}
\begin{align}\label{41}
\left.\frac{\partial \log P_g}{\partial \mu_0}\right|_{f} = \frac{6}{5} \, \frac{\log D_{\Lambda} \, (1+\beta_{\Lambda}\,\hat{\mu}^{2}) + \beta_{\Lambda}\,\hat{\mu}^{2}}{1+\beta_{\Lambda}\,\hat{\mu}^{2}},
\end{align}
\begin{eqnarray}\label{42}
\left.\frac{\partial \log P_g}{\partial \mu_2}\right|_{f} = \left.\frac{\partial \log P_g}{\partial \mu_0}\right|_{f} \, x^2,
\end{eqnarray}
\begin{eqnarray}\label{43}
\left.\frac{\partial \log P_g}{\partial \mu_4}\right|_{f} = \left.\frac{\partial \log P_g}{\partial \mu_0}\right|_{f} \, x^4.
\end{eqnarray}
As we can see, the only angular dependence appear in the derivatives respect to $\mu_2$ and $\mu_4$ which involve even powers of $x$. Thus, we can extract this dependence and define the following function,
\begin{widetext}
\begin{align}\label{44}
f^{x}_{i j} (\hat{\mu}, \alpha, \phi) = \int_0^{2 \pi} \, d\varphi \, \int_{0}^{\arccos(1-2f_{sky})} \, \sin \theta \, d\theta \, \int_0^{2 \pi} \, d\rho \,\, &(\delta_{1 i} + \delta_{2 i} + \delta_{3 i} + x^2 \, \delta_{4 i} + x^4 \, \delta_{5 i}) \nonumber \\ 
&(\delta_{1 j} + \delta_{2 j} + \delta_{3 j} + x^2 \, \delta_{4 j} + x^4 \, \delta_{5 j}),
\end{align}
\end{widetext}
where $x=x(\alpha, \phi, \hat{\mu}, \rho, \theta, \varphi)$ and $i, j = E, D_{\Lambda}, \mu_0, \mu_2, \mu_4 = 1, 2, 3, 4, 5$. Notice that for $i, j = 1, 2, 3$ we have $f^{x}_{i j} = 8 \pi^2 \, f_{sky}$, and we recover the isotropic case for the Fisher matrix. Finally, the Fisher matrix for the redshift-space power spectrum in the presence of a preferred direction pointing in the $(\alpha,\phi)$ direction can be written as,
\begin{widetext}
\begin{equation}\label{45}
F_{i j} (z_a, \alpha, \phi) = \frac{V_z (z_a)}{16 \, \pi^3} \, \int_{-1}^{1} \, d\hat{\mu} \, \int_{k_{min}}^{k_{max}} \, k^2 \, \left.\frac{\partial \log P_g (z_a, \hat{\mu}, k)}{\partial p_{i}}\right|_{f} \, \left.\frac{\partial \log P_g (z_a, \hat{\mu}, k)}{\partial p_{j}}\right|_{f} \, f^{x}_{i j} (\hat{\mu}, \alpha, \phi) \, \left. \left[ \frac{\bar{n} \hat{P_g} (z_a, \hat{\mu}, k)}{1+\bar{n} \hat{P_g} (z_a, \hat{\mu}, k)} \right] \right|_{f} \, dk,
\end{equation}
\end{widetext}
being $\hat{P_g} = P_g \, e^{-k^{2} \, \hat{\mu}^{2} \, \sigma_{r}^{2}}$ and in this expression,
\begin{equation}\label{46}
\left.\frac{\partial \log P_g}{\partial p_{4}}\right|_{f} = \left.\frac{\partial \log P_g}{\partial p_{5}}\right|_{f} = \left.\frac{\partial \log P_g}{\partial \mu_0}\right|_{f}.
\end{equation}
The values for $k_{min}$ and $k_{max}$ are the same as in the previous section. 

\begin{figure} %[h!]
  	\includegraphics[width=0.4975\textwidth]{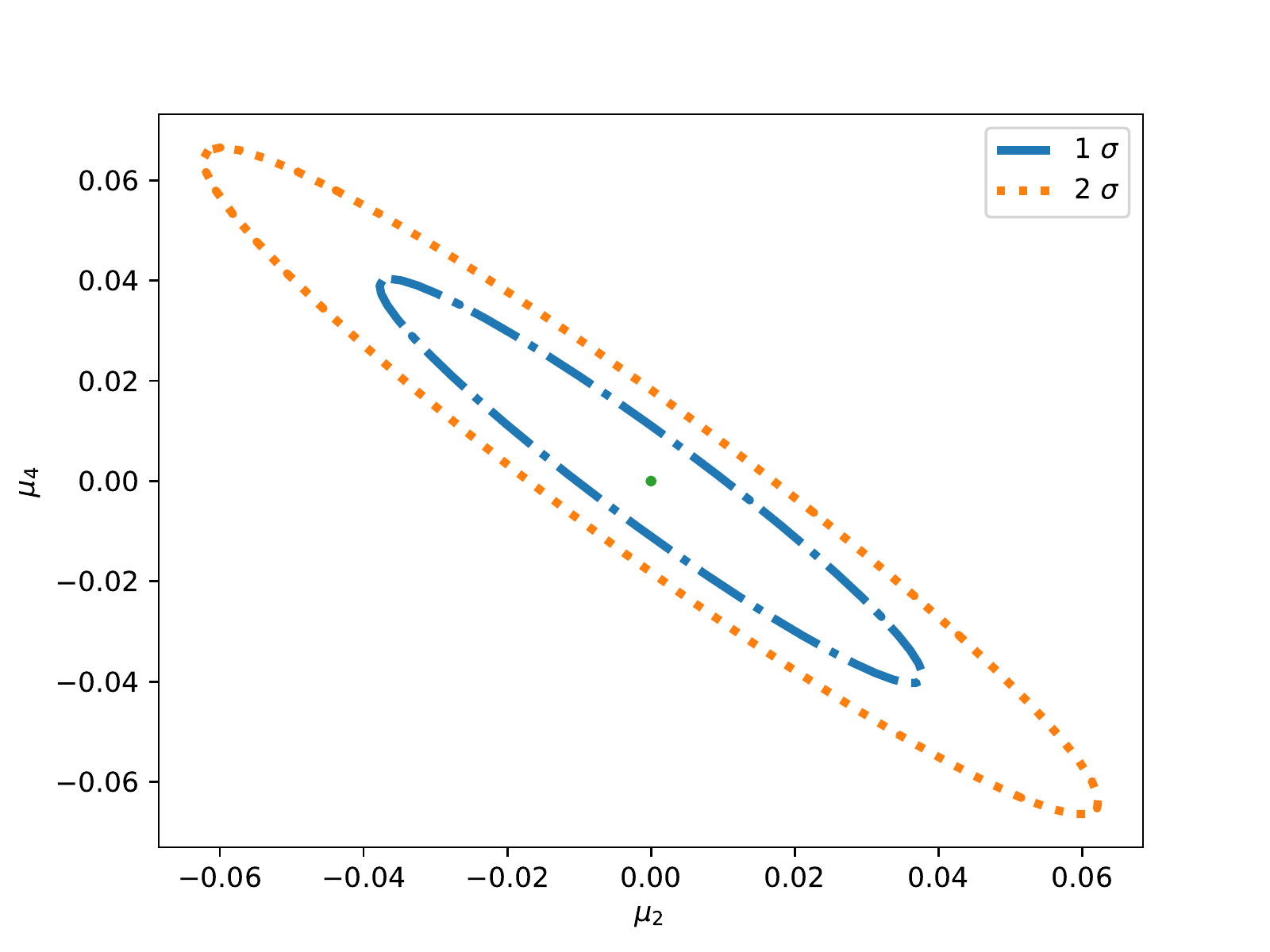}
		\caption{\footnotesize{Marginalized 1$\sigma$ and 2$\sigma$ regions for $\mu_2$ and $\mu_4$ using the information of the redshift space power spectrum and considering $\alpha = 0$ for an Euclid-like survey}}
  \label{Figure_4}
\end{figure}

Notice that the final Fisher matrix (\ref{45}) depends on the angles $(\alpha, \phi)$. We could have considered them as additional cosmological parameters $p_{i}$ and obtain and extended Fisher matrix. However, since we are  considering an isotropic fiducial model, the corresponding entries would be identically zero. Instead, we
will study that dependence of the errors on the 
orientation of the vector $\hat A$. Thus, we find that errors are maximized for $\alpha = 0$, i.e. when the preferred direction points towards the center of the survey, for any value of $\phi$, whereas they are minimized for $\alpha = \pi / 2$ for any value of $\phi$. Notice that in any case errors vary at most in a $10 \%$ of their values. We use the same fiducial cosmology as in the previous section for and Euclid-like survey. Results are summarized in Table \ref{Table3} and in Fig. \ref{Figure_4} we plot the 1-$\sigma$ and 2-$\sigma$ contours for $\mu_2$ and $\mu_4$ summing all the information in each bin. 

\begin{table}[htbp]
\begin{tabular}{|c|c|c|c|c|c|}
\hline
$z$ & $k_{max}$ & $n \, \times \, 10^{-3}$ & $\delta \mu_0 / \mu_0 (\%)$ & $\delta \mu_2 / \mu_0 (\%)$ & $\delta \mu_4 / \mu_0 (\%)$ \\
\hline \hline
0.6  &  0.195  &  3.56  &  2.35  &  10.1  &  10.7 \\ \hline
0.8  &  0.225  &  2.42  &  1.70  &  7.28  &  7.78 \\ \hline
1.0  &  0.260  &  1.81  &  1.31  &  5.61  &  6.00 \\ \hline
1.2  &  0.299  &  1.44  &  1.30  &  5.56  &  5.94 \\ \hline
1.4  &  0.343  &  0.99  &  1.29  &  5.51  &  5.89 \\ \hline
1.8  &  0.447  &  0.33  &  1.27  &  5.42  &  5.78 \\ \hline
\end{tabular}
\caption{Redshift bins, $k_{max}$ values in $\textrm{h/Mpc}$ units, galaxy densities in $(\textrm{h/Mpc})^{3}$ units and relative errors for $\mu_0$, $\mu_2$ and $\mu_4$ for an Euclid-like survey using the  redshift space power spectrum with $\alpha = 0$. We compare $\delta \mu_2$ and $\delta \mu_4$ with respect to $\mu_0$ because their fiducial values are zero.}
\label{Table3}
\end{table}

As we can see, with this method we obtain slightly larger errors for $\mu_0$, $\mu_2$ and $\mu_4$ than in the previous section.  \\

%%%%%%%%%%%%%%%%%%%%%%%%%%%%%%%%%%%%%%%%%%%%%%%%%%%%%%%%%%%%%%%%%%%%%%%%%%%%%%%%%%%%
\section{Fisher analysis for the weak lensing power spectrum}\label{sec9}
%%%%%%%%%%%%%%%%%%%%%%%%%%%%%%%%%%%%%%%%%%%%%%%%%%%%%%%%%%%%%%%%%%%%%%%%%%%%%%%%%%%%

In this section we extend in a simple way the Fisher matrix formalims for the weak lensing convergence power spectrum in the presence of a preferred direction. To do that, we have to analyze the multipole power spectrum for the convergence (\ref{s1.99}) and sum over all the multipoles $r$ and $\vec{\ell}$. The Fisher matrix is of the following form,
\begin{align}\label{s1.103b}
F_{\alpha \beta} = \sum_{\vec{\ell}_a, \vec{\ell}_b} \, \sum_{r r'} \, \frac{\partial P_{\kappa \,\, i j}^{\,r}(\vec{\ell}_a)}{\partial p_{\alpha}} \, \left[  \mathrm{Cov}_{j m n i}^{r r' } (\vec{\ell}_a, \vec{\ell}_b) \right]^{-1} \, \frac{\partial P_{\kappa \,\, m n}^{\,r'}(\vec{\ell}_b)}{\partial p_{\beta}},
\end{align}
where we are summing in indexes $i$, $j$, $m$ and $n$. We obtain the covariance matrix as an extension of the covariance matrix in the isotropic space in Appendix B. The corresponding Fisher matrix reads,
\begin{widetext}
\begin{align}\label{s1.108a}
F_{\alpha \beta} = f_{sky} \, \int_{-1}^{1} \, \frac{d \Upsilon}{\pi \, \sqrt{1-\Upsilon^2}} \, \sum_{r} \, \sum_{\ell} \, \Delta \ln \ell \, \frac{(2 \ell + 1) \, \ell}{2 \, (2r+1)} \, \left.\frac{\partial P_{\kappa \,\, i j}^{\,r}}{\partial p_{\alpha}}\right|_{f} \, C_{j m}^{-1} \, \left.\frac{\partial P_{\kappa \,\, m n}^{\,r}}{\partial p_{\beta}}\right|_{f} \, C_{n i}^{-1},
\end{align}
\end{widetext}
where the sub-index $f$ denotes that the corresponding quantity is evaluated on the fiducial model and,
\begin{align}\label{s1.108b}
C_{i j} = P_{\kappa \,\, i j} + \frac{\gamma_{int}^2}{\hat{n}_i} \, \delta_{i j},
\end{align}
being $\gamma_{int} = 0.22$ the intrinsic ellipticity \cite{Hilbert:2016ylf}, $\hat{n}_{i}$ the galaxies per steradian in the $i$-th bin,
\begin{equation}\label{s1.105}
\hat{n}_{i}=n_{\theta} \, \frac{\int_{\bar{z}_{i-1}}^{\bar{z}_{i}} n(z) \, dz}{\int_{0}^{\infty} n(z) \, dz},
\end{equation}
where $n_{\theta}$ is the areal galaxy density. We sum in $\ell$ with $\Delta \ln \ell = 0.1$ from $\ell_{min} = 5$ to $\ell_{max} = \chi (z_{\alpha'}) \, k_{max}$ being $\alpha' = \mathrm{min}(\alpha, \beta)$. For the multipole power spectrum we use the following expression,
\begin{equation}\label{s1.105b}
P_{\kappa \,\, i j}^{\,r} (\ell, \Upsilon) = \frac{1}{4}\sum_a P_{i j} (z_a,\ell)\;M^r_{\alpha} (\Upsilon) .
\end{equation}
where
\begin{align}\label{s1.114}
P_{i j} (z_a,\ell) = \frac{9 \, H_0^3 \, \Omega_m^2}{4}& \, \frac{(1+z_a)^2}{E_a} \Delta z_a \nonumber \\
 &g_i (z_a) \, g_j (z_a) \, D_a^2 \, P\left(\frac{\ell}{\chi (z_a)}\right),
\end{align}

Regarding the parameters $\theta_{\alpha}$, it can be proved that in each bin, the power spectrum depends on four independent parameters, which are chosen as ($E_a$, $\gamma_a$, $\mu_{Q \,\, a }$, $\mu_{h \,\, a }$) where the sub-index $a$ denotes different redshift bins, so that we have a total Fisher matrix of size $4n \times 4n$, being $n$ the total number of $z$ bins. 

\begin{figure} %[h!]
  	\includegraphics[width=0.4975\textwidth]{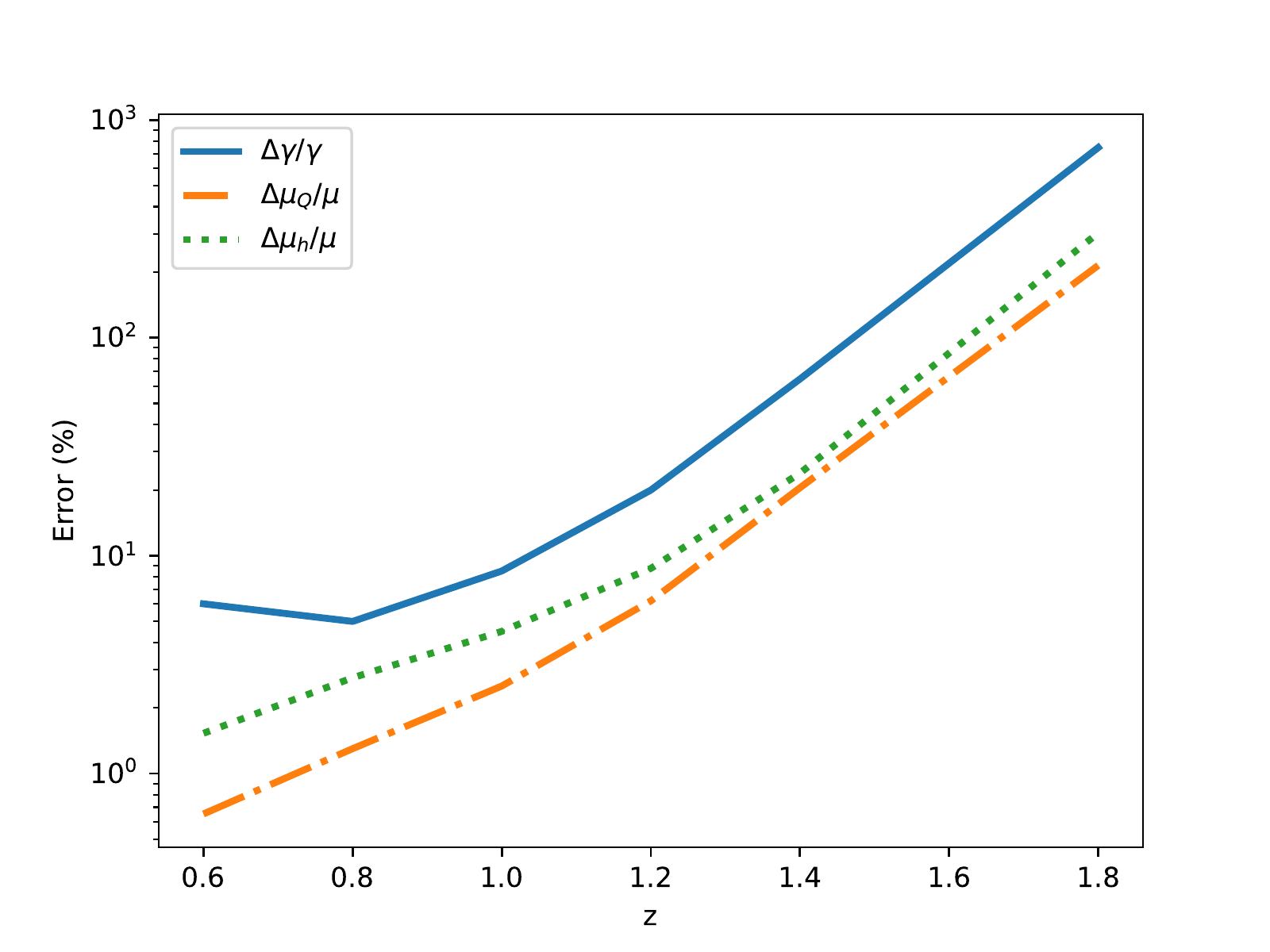}
		\caption{\footnotesize{Relative errors for $\gamma$, $\mu_{Q}$ and $\mu_{h}$ using weak lensing information for an Euclid-like survey.}}
  \label{Figure_5}
\end{figure}

For the sake of simplicity, we will consider that the modified gravity parameters are isotropic and scale invariant, i.e. $\gamma_a=\gamma(z_a)$, 
$\mu_{Q\, a}=\mu_Q(z_a)$ and $\mu_{h\, a}=\mu_h(z_a)$.  The
(non-vanishing) derivatives, which are evaluated in $\Lambda$CDM as  fiducial model, are,

\begin{align}\label{s1.109}
\left.\frac{\partial P_{\kappa \,\, i j}^{\,0}}{\partial \gamma_a}\right|_{f} = P_{i j} (z_a),
\end{align}
\begin{align}\label{s1.110}
\left.\frac{\partial P_{\kappa \,\, i j}^{\,1}}{\partial \mu_{Q \,\, a }}\right|_{f} = -4 \, P_{i j} (z_a),
\end{align}
\begin{align}\label{s1.111}
\left.\frac{\partial P_{\kappa \,\, i j}^{\,0}}{\partial \mu_{h \,\, a }}\right|_{f} = \frac{2}{3} \, (2-\Upsilon^2) \, P_{i j} (z_a),
\end{align}
\begin{align}\label{s1.112}
\left.\frac{\partial P_{\kappa \,\, i j}^{\,2}}{\partial \mu_{h \,\, a }}\right|_{f} = \frac{4}{3} \, (2-\Upsilon^2) \, P_{i j} (z_a),
\end{align}
\begin{align}\label{s1.113}
\left.\frac{\partial P_{\kappa \,\, i j}^{\,0}}{\partial E_a}\right|_{f} = - \frac{P_{i j} (z_a)}{E_a}& + \sum_b \, \frac{1}{g_i (z_b)} \frac{\partial g_i (z_b)}{\partial E_a} \, P_{i j} (z_b) \nonumber\\
&+ \sum_b \, \frac{1}{g_j (z_b)} \frac{\partial g_j (z_b)}{\partial E_a} \, P_{i j} (z_b),
\end{align}
with,
\begin{eqnarray}\label{s1.115a}
\frac{\partial g_{i} (z_{b})}{\partial E_{a}}=\frac{\Delta z_{a}}{H_0 \, E_{a}^{2}} \, \left[ - \hat{\theta} (z_{a}-z_{b}) \, \chi (z_{b}) \, \int_{z_{a}}^{\infty} \frac{n_{i} (z')}{\chi (z')^{2}} \, dz' \right. \nonumber\\
\left. + \, \theta (z_{b}-z_{a}) \, \int_{z_{b}}^{\infty} \left(1-\frac{\chi (z_{b})}{\chi (z')} \right) \frac{n_{i} (z')}{\chi (z')} \, dz' \right],\nonumber\\
\end{eqnarray}
where we have discretized the integration of $E(z)^{-1}$ in $\chi (z)$ for the different bins, and the step functions $\theta(z)$ and $\hat\theta(x)$ are defined so that $\hat{\theta}(0)=0$ and $\theta(0)=1$.  We consider Euclid as a weak lensing survey with a fraction of the sky $f_{sky} =0.364$, $z_{mean}=0.9$ and $n_{\theta}=35$ galaxies per square arc minute with $\delta z=0.05$. We summarize the results in Table \ref{Table4} and in Fig. \ref{Figure_5}. 

\begin{figure} %[h!]
  	\includegraphics[width=0.4975\textwidth]{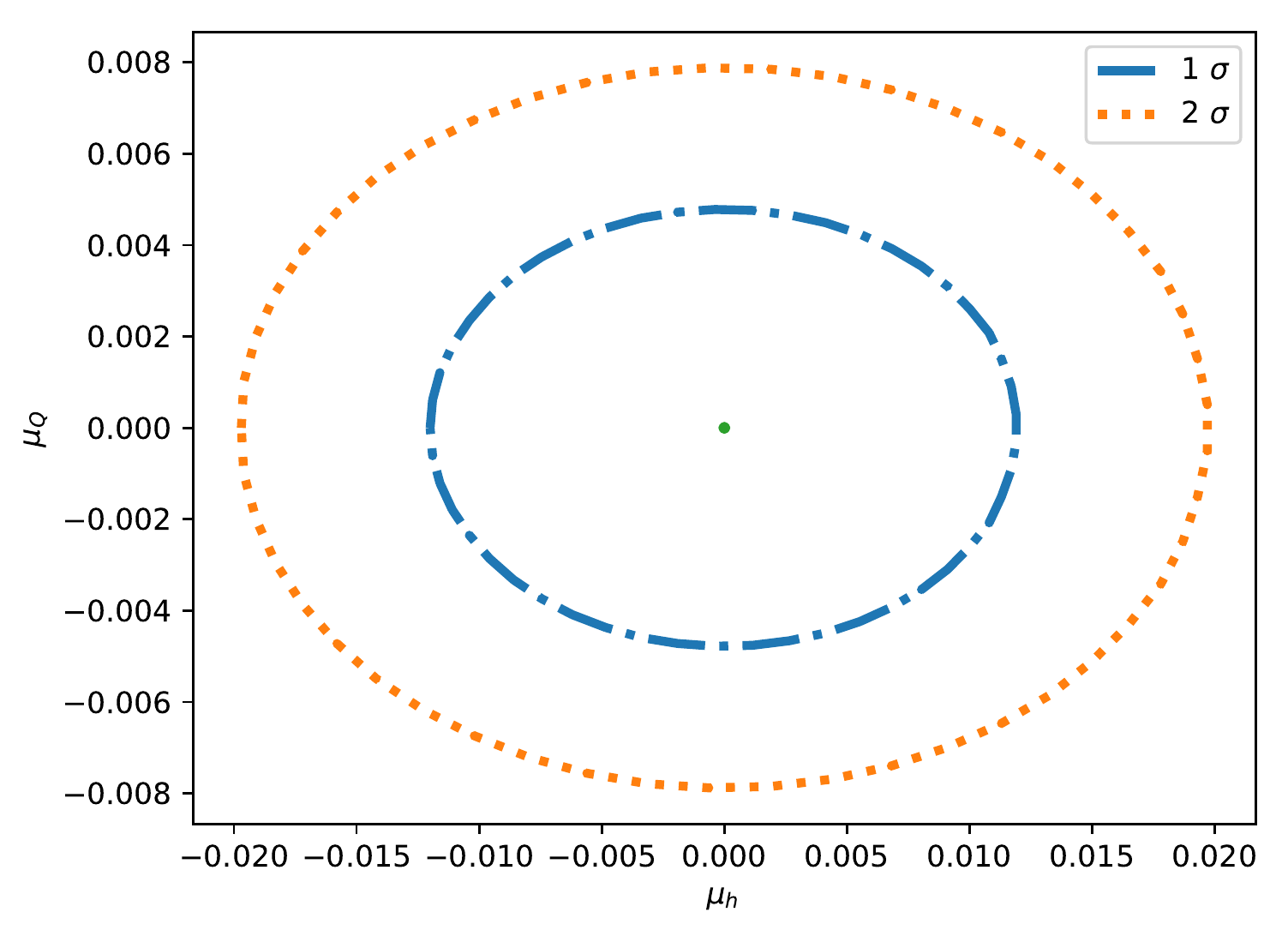}
		\caption{\footnotesize{Marginalized  1$\sigma$ and 2$\sigma$  regions for for $\mu_{Q}$ and $\mu_{h}$ summing the information of the whole redshift range for an Euclid-like survey.}}
  \label{Figure_6}
\end{figure}

Finally, if we further assume that $\gamma$, $\mu_{Q}$ and $\mu_{h}$ are just constants, we can sum the information in all redshift bins. We plot the corresponding 1-$\sigma$ and 2-$\sigma$ contours for $\mu_{Q}$ and $\mu_{h}$ in Fig. \ref{Figure_6}.

\begin{table}[htbp]
\begin{tabular}{|c|c|c|c|c|}
\hline
$z$ & $\ell_{max}$ & $\delta \gamma / \gamma (\%)$ & $\delta \mu_{Q} / \mu (\%)$ & $\delta \mu_{h} / \mu (\%)$ \\
\hline \hline
0.6  &  300   &  6.01  &  0.65  &  1.53  \\ \hline
0.8  &  438   &  4.99  &  1.30  &  2.75  \\ \hline
1.0  &  598   &  8.50  &  2.52  &  4.49  \\ \hline
1.2  &  783   &  20.0  &  6.20  &  8.74  \\ \hline
1.4  &  996   &  64.7  &  20.5  &  23.9  \\ \hline
1.8  &  1520  &  747   &  215   &  303   \\ \hline
\end{tabular}
\caption{Redshift bins,  $\ell_{max}$ values and relative errors for $\gamma$, $\mu_{Q}$ and $\mu_{h}$ for the Euclid forecast. We compare $\mu_{Q}$ and $\mu_{h}$ with $\mu$ because the fiducial values of $\mu_{Q}$ and $\mu_{h}$ are zero.}
\label{Table4}
\end{table}

We can see that lensing convergence measurements are very  sensitive to the dipole term $P^1_{\kappa\, ij}$, so that  errors in $\mu_{Q}$ are much smaller than for $\gamma$ and $\mu_{h}$. Notice also that multipoles $r=3,4$ do not appear in the derivatives (\ref{s1.109})-(\ref{s1.113}) since those terms are quadratic in $\mu_{Q}$ and $\mu_{h}$ so that
on the fiducial $\Lambda$CDM cosmology the corresponding derivatives vanish.     For the same reason, the Fisher matrix for the rotation power spectrum also 
 vanishes.

%%%%%%%%%%%%%%%%%%%%%%%%%%%%%%%%%%%%%%%%%%%%%%%%%%%%%%%%%%%%%%%%%%%%%%%%%%%%%%%%%%%%
\section{Forecasting primordial anisotropies}\label{sec10}
%%%%%%%%%%%%%%%%%%%%%%%%%%%%%%%%%%%%%%%%%%%%%%%%%%%%%%%%%%%%%%%%%%%%%%%%%%%%%%%%%%%%

So far we have studied the effects of  preferred directions in the evolution of density and metric perturbations, but anisotropies could also be present in the primordial
curvature power spectrum \cite{Pullen,Ackerman:2007nb}. In this case, and assuming parity symmetry, the leading effects can 
be described by a modification of the primordial power spectrum from $P(k)$ to $P'(\vec{k})$ such that,
\begin{align}\label{s8b.1}
P'\left(\vec{k}\right) = \left(1+g_* \, x^2\right) \, P(k). 
\end{align}
Assuming a scale-independent modification, $g_*$ is just a dimensionless constant. We can use the Fisher formalism of
Sec. \ref{sec3} and Sec. \ref{sec8} to forecast the sensitivity
with which future surveys could measure the $g_*$ parameter. With that purpose, we consider both, the multipole power spectrum for the matter distribution and the multipole power spectrum for lensing convergence. For
clustering  we consider the following independent parameters in the Fisher analysis ($\beta_a$, $D_a$, $g_{* \,\, a}$, $b_a$), whereas for lensing we take ($E_a$, $L_a$, $g_{* \,\, a}$) where the sub-index $a$ denotes  the different redshift bins and $L_a \equiv \Omega_m^2 D_a^2 \sigma_8^2$. We summarize the results in Table \ref{Table5} for an Euclid-like survey as in previous sections.

\begin{table}[htbp]
\begin{tabular}{|c|c|c|c|c|c|}
\hline
$z$ & $k_{max}$ & $\ell_{max}$ & $n \times 10^{-3}$ & $100\times\delta g_*^{C}$ & $100\times\delta g_*^{L}$ \\
\hline \hline
0.6  & 0.195 &  300   &  3.56  &  0.61  & 4.52   \\ \hline
0.8  & 0.225 &  438   &  2.42  &  0.43  & 6.84   \\ \hline
1.0  & 0.260 &  598   &  1.81  &  0.32  & 9.24   \\ \hline
1.2  & 0.299 &  783   &  1.44  &  0.30  & 15.3   \\ \hline
1.4  & 0.343 &  996   &  0.99  &  0.29  & 37.2   \\ \hline
1.8  & 0.447 &  1520  &  0.33  &  0.29  & 566    \\ \hline
\end{tabular}
\caption{Redshift bins, $k_{max}$ values in $\textrm{h/Mpc}$ units, $\ell_{max}$ values, galaxy densities in $(\textrm{h/Mpc})^{3}$ units  and forecasted absolute errors for $g_*$ from clustering (C) and lensing (L) for an Euclid-like survey.}
\label{Table5}
\end{table}

As we can see, we have better precision with the multipole power spectrum of galaxy distribution. If we sum the information of clustering and lensing and in each bin, we obtain and absolute error $\delta g_* = 1.4\times 10^{-3}$.

%%%%%%%%%%%%%%%%%%%%%%%%%%%%%%%%%%%%%%%%%%%%%%%%%%%%%%%%%%%%%%%%%%%%%%%%%%%%%%%%%%%%
\section{Conclusions}\label{sec11}
%%%%%%%%%%%%%%%%%%%%%%%%%%%%%%%%%%%%%%%%%%%%%%%%%%%%%%%%%%%%%%%%%%%%%%%%%%%%%%%%%%%%

In this work we have considered possible observational signatures of  modifications of  gravity  involving preferred
spatial directions. In the model-independent approach, these theories 
can be parametrized in the sub-Hubble regime and using the quasi-static approximation by four effective parameters $\mu(a,k,x)$, $\gamma(a,k,x)$, $\mu_Q(a,k,x)$ and $\mu_h(a,k,x)$. We have analyzed the
effects of the existence of a preferred direction in galaxy distribution and weak lensing observations. In the galaxy power spectrum, we find that a new angular dependence
on the line of sight appears which is different from the usual effect induced by redshift space distortions. In the lensing case, we have two new effects. On one hand,
 a dependence on the line of sight is introduced in the shear and convergence power spectra which is absent in the isotropic case. In particular the $\mu_Q$ parameter introduces
a dipole contribution, whereas the $\mu_h$ produces a quadrupole term. 
On the other hand, images rotation is induced in addition to the standard convergence and  shear effects. Thus, we have found a useful relation between the
different power spectra,
\begin{align}\label{s1.115b}
P_{\gamma_1} + P_{\gamma_2} = P_{\kappa} + P_{\omega},
\end{align}
which shows that even though $P_{\omega}$ cannot be measured directly using weak lensing maps, it can be derived from $P_{\gamma_1}$, $P_{\gamma_2}$ and $P_{\kappa}$.

We have also forecasted the precision with which future surveys will be 
able to measure the four effective parameters. With that purpose 
we have extended the standard Fisher matrix approach in order
to include the presence of preferred directions. In particular,  explicit expressions for the covariance matrices for the multipole galaxy power spectrum and for the convergence power spectrum have been derived in the Appendices. For the galaxy power spectrum, we have considered two different 
approaches. On one hand, we have obtained the Fisher matrix for the 
power spectrum in redshift space, which allows us to include the  Alcock-Paczynski effect. On the other, we computed the corresponding Fisher matrices for the multipole power spectra. In both cases, we obtain that 
the precision on measurements of the effective Newton constant $\mu=\mu_0+\mu_2x^2+\mu_4 x^4$  for an Euclid-like survey will be around $1\%$ for $\mu_{0}$ and a few percent for $\mu_2$ and $\mu_4$. Very much as in the 
clustering case, the lensing forecast indicates that the $\gamma$ parameter could be measured with a few percent precision, wheras  $\mu_Q$ and $\mu_h$ parameters could be determined with precision around $1 \%$.

\vspace{0.2cm}
{\bf Acknowledgements:}
M.A.R acknowledges support from UCM predoctoral grant.  This work has been supported by the MINECO (Spain) project FIS2016-78859-P(AEI/FEDER, UE).

%%%%%%%%%%%%%%%%%%%%%%%%%%%%%%%%%%%%%%%%%%%%%%

%%%%%%%%%%%%%%%%%%%%%%%%%%%%%%%%%%%%%%%%%%%%%%%%%%%%%%%%%%%%%%%%%%%%%%%%%%%%%%%%%%%%
\section{Appendix A}\label{secA1}
%%%%%%%%%%%%%%%%%%%%%%%%%%%%%%%%%%%%%%%%%%%%%%%%%%%%%%%%%%%%%%%%%%%%%%%%%%%%%%%%%%%%

In this appendix we will calculate the covariance matrix for an anisotropic matter power spectrum. To do that, we start by reviewing the standard calculation in the 
isotropic case \cite{Yamamoto:2005dz, Takada:2007fq, Kayo:2013aha, Guzik} and then we will extend it 
to include a preferred direction. Let us thus start with the estimator for the matter power spectrum in the isotropic case,
\begin{align}\label{A.1}
\hat{P}(k_i) = V_f \, \int_{k_i} \frac{d^3 \vec{k}}{V_s (k_i)} \, \delta(\vec{k}) \delta(-\vec{k}),
\end{align}
where $V_f = (2 \pi)^3 / V$, being $V$ the volume of the survey and $\int_{k_i} d^3 \vec{k} = V_s (k_i) = 4 \pi k_i^2 dk_i$ the volume of the $k_i$ bin. Thus if we use,

\begin{align}\label{A.2}
\langle \delta(\vec{k}_1) \delta(\vec{k}_2) \rangle = \delta_D (\vec{k}_1 + \vec{k}_2) \, P(\vec{k}_1),
\end{align}
we can prove that $\langle \hat{P}(k_i) \rangle = P(k_i)$, where $\delta_D (0) = 1/V_f$. Now, we want to calculate the covariance matrix, defined as
\begin{align}\label{A.3}
C (k_i, k_j) = \langle \hat{P}(k_i) \hat{P}(k_j) \rangle - P(k_i) P(k_j).
\end{align}
We consider only the gaussian case, so that
\begin{align}\label{A.4}
\langle F(s_1) F(s_2) F(s_3) F(s_4) \rangle \, &= \,  \langle F(s_1) F(s_2) \rangle \langle F(s_3) F(s_4) \rangle \nonumber\\ 
& + \langle F(s_1) F(s_3) \rangle \langle F(s_2) F(s_4) \rangle \nonumber\\
& + \langle F(s_1) F(s_4) \rangle \langle F(s_2) F(s_3) \rangle.
\end{align}
Using this relation we obtain,
\begin{align}\label{A.5}
C (k_i, k_j) \, = \, & \frac{V_f^2 \, \delta_D (0)}{V_s (k_i) V_s (k_j)} \, \int_{k_i} d^3 \vec{k} \, \int_{k_j} d^3 \vec{k'} \nonumber\\ &[\delta_D(\vec{k}+\vec{k'})+\delta_D(\vec{k}-\vec{k'})] \, P(k)^2,
\end{align}
where we have used the property $\delta^2_D(x) = \delta_D (0) \, \delta_D (x)$, and $P(\vec{k}) = P(k)$ being $k = |\vec{k}|$. We consider that $P(k)$ is constant in the integral of the $k_i$-bin, so
that we can extract it from the integral as $P(k_i)$. Thus, the integrals in $k$ and $k'$ become,

\begin{align}\label{A.6}
\int_{k_i} d^3 \vec{k} \, \int_{k_j} d^3 \vec{k'} \, [\delta_D(\vec{k}+\vec{k'})+\delta_D(\vec{k}-\vec{k'})] = 2 \, V_s (k_j) \, \delta_{i j}.
\end{align}
Finally if we take into account the shot noise, the observable galaxy density contrast becomes, 
\begin{align}\label{A.2b}
\delta^{obs}(k) = \delta(k) + \epsilon(k),
\end{align}
where $\epsilon (k)$ is a random gaussian variable with $\langle \epsilon (k) \rangle = 0$ and $\langle \epsilon (k) \epsilon (k') \rangle = \delta_D (k - k')/\bar{n}$
with $\bar n$ the average galaxy number density. Then the observable power spectrum becomes $P^{obs}(k_i) = P(k_i) + \frac{1}{\bar{n}}$ and we obtain,

\begin{align}\label{A.7}
C (k_i, k_j) = \delta_{i j} \, \frac{2 V_f}{V_s (k_i)} \, \left[ P(k_i) + \frac{1}{\bar{n}} \right]^2.
\end{align}
Now we want to extend this procedure for the case in which we have an anisotropic power spectrum depending not only on the full wavevector $\vec k$, but also on its orientation with respect to the line of sight $\hat{n}$. In particular, we have a power spectrum $P(\vec{k},\hat\mu)$ where $\hat\mu =\hat k \cdot \hat n$. We can decompose this power spectrum in the form,
\begin{align}\label{A.8}
P(\vec{k},\hat\mu) = \sum_{\ell} \, P_{\ell} (\vec{k}) \, \mathcal{L}_{\ell} (\hat\mu),
\end{align}
being $\mathcal{L}_{\ell} (\hat\mu)$ the Legendre polynomials,
\begin{align}\label{A.9}
P_{\ell} (\vec{k}) = \frac{2\ell+1}{2} \, \int_{-1}^1 d\hat\mu \, P(\vec{k},\hat\mu) \, \mathcal{L}_{\ell} (\hat\mu).
\end{align}
We define the estimator for this multipole power spectrum in the following way,
\begin{align}\label{A.10}
\hat{P}_{\ell} (\vec{k}_i) = V_f \, \int_{\vec{k}_i} \frac{d^3 \vec{k}}{V_s (\vec{k}_i)} \, \frac{2\ell+1}{2} \, \int_{-1}^1 d\hat\mu \, \delta(\vec{k},\hat\mu) \delta(-\vec{k},\hat\mu) \, \mathcal{L}_{\ell} (\hat\mu),
\end{align}
where in general $\int_{\vec{k}_i} d^3 \vec{k} = V_s (\vec{k}_i) = k_i^2 \, dk_i \, dx_i \, d\phi_i$, being $x_i = \cos \theta_i$. If we consider,
\begin{align}\label{A.11}
\langle \delta(\vec{k}_1, \hat\mu) \delta(\vec{k}_2, \hat\mu) \rangle = \delta_D (\vec{k}_1 + \vec{k}_2) \, P(\vec{k}_1, \hat\mu),
\end{align}
we can prove that $\langle \hat{P}_{\ell} (\vec{k}_i) \rangle = P_{\ell} (\vec{k}_i)$. With this estimator we can calculate the covariance matrix,
\begin{align}\label{A.12}
C_{\ell \ell'} (\vec{k}_i, \vec{k}_j) = \langle \hat{P}_{\ell}(\vec {k}_i) \hat{P}_{\ell'}(\vec{k}_j) \rangle - P_{\ell}(\vec{k}_i) P_{\ell'}(\vec{k}_j).
\end{align}
As in the isotropic case, we consider only  gaussian perturbations satisfying (\ref{A.4}), so that
\begin{align}\label{A.13}
C_{\ell \ell'} (\vec{k}_i, \vec{k}_j) &= V_f^2 \, \frac{(2\ell+1)(2\ell'+1)}{4} \, \int_{\vec{k}_i} \frac{d^3 \vec{k}}{V_s (\vec{k}_i)} \, \int_{\vec{k}_j} \frac{d^3 \vec{k'}}{V_s (\vec{k}_j)} \nonumber\\
& \int_{-1}^1 d\hat\mu \, \int_{-1}^1 d\hat\mu' \mathcal{L}_{\ell} (\hat\mu) \, \mathcal{L}_{\ell'} (\hat\mu') \nonumber\\
& \left[ \langle \delta(\vec{k}, \hat\mu) \delta(\vec{k}', \hat\mu') \rangle \, \langle \delta(-\vec{k}, \hat\mu) \delta(-\vec{k}', \hat\mu') \rangle \right. \nonumber\\
& + \left. \langle \delta(\vec{k}, \hat\mu) \delta(-\vec{k}', \hat\mu') \rangle \, \langle \delta(-\vec{k}, \hat\mu) \delta(\vec{k}', \hat\mu') \rangle \right].
\end{align}

At this stage we use the distant observer approximation in which we assume that the integrand of (\ref{A.13}) is non negligible only when $\hat\mu' \simeq \hat\mu$, then we obtain,
\begin{align}\label{A.14}
C_{\ell \ell'} (\vec{k}_i, \vec{k}_j) & \simeq V_f^2 \, \frac{(2\ell+1)(2\ell'+1)}{2} \, \int_{\vec{k}_i} \frac{d^3 \vec{k}}{V_s (\vec{k}_i)} \, \int_{\vec{k}_j} \frac{d^3 \vec{k'}}{V_s (\vec{k}_j)} \nonumber\\
& \int_{-1}^1 d\hat\mu \, \mathcal{L}_{\ell} (\hat\mu) \, \mathcal{L}_{\ell'} (\hat\mu) \nonumber\\
& \left[ \langle \delta(\vec{k}, \hat\mu) \delta(\vec{k}', \hat\mu) \rangle \, \langle \delta(-\vec{k}, \hat\mu) \delta(-\vec{k}', \hat\mu) \rangle \right. \nonumber\\
& + \left. \langle \delta(\vec{k}, \hat\mu) \delta(-\vec{k}', \hat\mu) \rangle \, \langle \delta(-\vec{k}, \hat\mu) \delta(\vec{k}', \hat\mu) \rangle \right].
\end{align}
Using (\ref{A.11}) and taking into account  once again that $\delta_D^2(x) = \delta_D (0) \delta_D (x)$ and $\delta_D (0) = 1/V_f$, we obtain,
\begin{align}\label{A1.15}
C_{\ell \ell'} & (\vec{k}_i, \vec{k}_j) \simeq V_f \, \frac{(2\ell+1)(2\ell'+1)}{2} \, \int_{\vec{k}_i} \frac{d^3 \vec{k}}{V_s (\vec{k}_i)} \, \int_{\vec{k}_j} \frac{d^3 \vec{k'}}{V_s (\vec{k}_j)} \nonumber\\
& \int_{-1}^1 d\hat\mu \, \mathcal{L}_{\ell} (\hat\mu) \, \mathcal{L}_{\ell'} (\hat\mu) \left[ \delta_D(\vec{k}+\vec{k'}) + \delta_D(\vec{k}-\vec{k'})  \right] \nonumber\\
& P(\vec{k}, \hat\mu) \, P(-\vec{k}, \hat\mu).
\end{align}
As done before, we consider that $P(\vec{k}, \hat\mu) \simeq P(\vec{k}_i, \hat\mu)$ in the integral and also that
\begin{align}\label{A1.16}
\int_{\vec{k}_i} d^3 \vec{k} \, \int_{\vec{k}_j} d^3 \vec{k'} \,\, \delta_D(\vec{k} - \vec{k'}) = V_s (\vec{k}_j) \, \delta_{k_i, k_j} \, \delta_{x_i, x_j} \, \delta_{\phi_i, \phi_j},
\end{align}
and
\begin{align}\label{A1.17}
\int_{\vec{k}_i} d^3 \vec{k} \, \int_{\vec{k}_j} d^3 \vec{k'} \,\, \delta_D(\vec{k} + \vec{k'}) = V_s (\vec{k}_j) \, \delta_{k_i, k_j} \, \delta_{x_i, -x_j} \, \delta_{\phi_i, \phi_j+\pi}.
\end{align}
Using these expressions we obtain,
\begin{widetext}
\begin{align}\label{A1.18}
C_{\ell \ell'} & (\vec{k}_i, \vec{k}_j) = \frac{V_f \, \delta_{k_i, k_j}}{V_s (\vec{k}_i)} \, [\delta_{x_i, x_j} \, \delta_{\phi_i, \phi_j}+\delta_{x_i, -x_j} \, \delta_{\phi_i, \phi_j+\pi}] \, \frac{(2\ell+1)(2\ell'+1)}{2} \int_{-1}^1 d\hat\mu \, \mathcal{L}_{\ell} (\hat\mu) \, \mathcal{L}_{\ell'} (\hat\mu) P(\vec{k}_i, \hat\mu) \, P(-\vec{k}_i, \hat\mu).
\end{align}

\end{widetext}
If we consider that the fiducial power spectrum is of the form $P(\vec{k}_i, \hat\mu) = P(k_i, \hat\mu)$, and we have only dependence on $x^2$ then $\delta_{x_i, -x_j} = \delta_{x_i, x_j}$. Also, we can integrate in $\phi_i$ so that $V_s (\vec{k}_i) = 2 \pi k_i^2 \, dk_i \, dx_i$, and
\begin{widetext}

\begin{align}\label{A1.19}
C_{\ell \ell'} & (\vec{k}_i, \vec{k}_j) = \delta_{\vec{k}_i, \vec{k}_j} \, \frac{2 \,V_f}{V_s (\vec{k}_i)} \, \frac{(2\ell+1)(2\ell'+1)}{2} \int_{-1}^1 d\hat\mu \, \mathcal{L}_{\ell} (\hat\mu) \, \mathcal{L}_{\ell'} \, (\hat\mu) P^2(k_i, \hat\mu).
\end{align}

\end{widetext}
Finally we take into account that the observable matter power spectrum is $P^{obs}(k_i, \hat\mu) = P(k_i, \hat\mu) + 1/\bar{n}$ so that the final expression for the covariance matrix we will 
consider reads
\begin{widetext}

\begin{align}\label{A.15}
C_{\ell \ell'} (\vec{k}_i, \vec{k}_j) =  \delta_{\vec{k}_i, \vec{k}_j} \, \frac{2 \,V_f}{V_s (\vec{k}_i)} \, \frac{(2\ell+1)(2\ell'+1)}{2} \int_{-1}^1 d\hat\mu \, \mathcal{L}_{\ell} (\hat\mu) \, \mathcal{L}_{\ell'} \, (\hat\mu) \left[P(k_i, \hat\mu) + \frac{1}{\bar{n}}\right]^2.
\end{align}

\end{widetext}

%%%%%%%%%%%%%%%%%%%%%%%%%%%%%%%%%%%%%%%%%%%%%%%%%%%%%%%%%%%%%%%%%%%%%%%%%%%%%%%%%%%%
\section{Appendix B}\label{secA2}
%%%%%%%%%%%%%%%%%%%%%%%%%%%%%%%%%%%%%%%%%%%%%%%%%%%%%%%%%%%%%%%%%%%%%%%%%%%%%%%%%%%%

In this appendix we will calculate the covariance matrix for an anisotropic lensing convergence power spectra. As in 
clustering case,  we start by reviewing \cite{Yamamoto:2005dz, Takada:2007fq, Kayo:2013aha, Guzik} the standard isotropic calculation. Let us thus first  obtain the estimator for the convergence power spectra in redshift bins $i,j$,
\begin{align}\label{A.16}
\hat{P}_{i j}(\ell_a) = A_f \, \int_{\ell_a} \frac{d^2 \vec{\ell}}{A_s (\ell_a)} \, \kappa_i(\vec{\ell}) \kappa_j(-\vec{\ell}),
\end{align}
where $A_f = (2 \pi)^2 / \Omega$, being $\Omega = 4\pi f_{sky}$ the total area of the survey and $\int_{\ell_a} d^2 \vec{\ell} = A_s (\ell_a) = 2 \pi \ell_a d\ell_a$
the area of the $\ell_a$ bin. Then if we use,
\begin{align}\label{A.17}
\langle \kappa_i(\vec{\ell}_1) \kappa_j(\vec{\ell}_2) \rangle = \delta_D (\vec{\ell}_1 + \vec{\ell}_2) \, P_{i j}(\vec{\ell}_1),
\end{align}
we can prove that $\langle \hat{P}_{i j}(\ell_a) \rangle = P_{i j}(\ell_a)$, where $\delta_D (0) = 1/A_f$. The corresponding covariance matrix is defined by,
\begin{align}\label{A.18}
C_{i j i' j'} (\ell_a, \ell_b) = \langle \hat{P}_{i j}(\ell_a) \hat{P}_{i' j'}(\ell_b) \rangle - P_{i j}(\ell_a) P_{i' j'}(\ell_b).
\end{align}
As before, we  take into account (\ref{A.4})  for gaussian perturbations  so that
\begin{align}\label{A.19}
&C_{i j i' j'} (\ell_a, \ell_b) = \frac{A_f^2 \, \delta_D (0)}{A_s (\ell_a) A_s (\ell_b)} \, \int_{\ell_a} d^2 \vec{\ell} \, \int_{\ell_b} d^2 \vec{\ell'} \nonumber\\ &[\delta_D(\vec{\ell}+\vec{\ell'}) P_{i i'}(\ell) P_{j j'}(\ell)+\delta_D(\vec{\ell}-\vec{\ell'}) P_{i j'}(\ell) P_{j i'}(\ell)],
\end{align}
where once more we have used the property $\delta_D^2(x) = \delta_D (0) \, \delta_D (x)$, and $P_{i j}(\vec{\ell}) = P_{i j}(\ell)$ being $\ell = |\vec{\ell}|$. By assuming that $P_{i j}(\ell)$ is constant within the $\ell_a$ bin,  we can extract it from the integral as $P_{i j}(\ell_a)$. Then, the integrals in $\ell$ and $\ell'$ become,
\begin{align}\label{A.20a}
\int_{\ell_a} d^2 \vec{\ell} \, \int_{\ell_b} d^2 \vec{\ell'} \,\, \delta_D(\vec{\ell} \pm \vec{\ell'}) = A_s (\ell_b) \, \delta_{a b}.
\end{align}
Finally if we take into account the intrinsic ellipticity $\gamma_{int}$, the observable convergence reads,
\begin{align}\label{A.22a}
\kappa^{obs}_{i}(\ell) = \kappa_{i}(\ell) + \gamma_{int} \, \epsilon_i (\ell),  
\end{align}
where $\epsilon_i (\ell)$ is a random gaussian variable with $\langle \epsilon_i (\ell) \rangle = 0$ and $\langle \epsilon_i (\ell) \epsilon_j (\ell') \rangle = \delta_D (\ell - \ell') \delta_{i j}/\hat{n}_i$, with $\hat n_i$
the areal galaxy density (per steradian) in the redshift bin $i$. Then the observable power spectrum is,
\begin{align}\label{A.21}
P^{obs}_{i j}(\ell) = P_{i j}(\ell) + \frac{\gamma_{int}^2}{\hat{n}_i} \delta_{i j}.
\end{align}
and the covariance is,
\begin{align}\label{A.20b}
&C_{i j i' j'} (\ell_a, \ell_b) = \frac{A_f \, \delta_{a b}}{A_s (\ell_a)} \, [P^{obs}_{i i'}(\ell_a) P^{obs}_{j j'}(\ell_a) + P^{obs}_{i j'}(\ell_a) P^{obs}_{j i'}(\ell_a)].
\end{align}
If we use the expressions for $A_f$ and $A_s(\ell_a)$ and consider $\ell \simeq (2\ell + 1)/2$ we obtain the known result \cite{Guzik},
\begin{widetext}
\begin{align}\label{A.22b}
&C_{i j i' j'} (\ell_a, \ell_b) = \frac{\delta_{a b}}{(2\ell_a + 1) f_{sky} d\ell_a} \, [P^{obs}_{i i'}(\ell_a) P^{obs}_{j j'}(\ell_a) + P^{obs}_{i j'}(\ell_a) P^{obs}_{j i'}(\ell_a)].
\end{align}
\end{widetext}
Now we want to obtain the covariance matrix for the case in which the power spectrum denpends not only on the full
$\vec\ell$ vector but also we have a dependence on the observation direction
$\hat n$, in particular a polar dependence ($\xi =\hat n\cdot \hat A$) where
$\hat A$ is the preferred direction.  As in (\ref{A.8}), we can  write
\begin{align}\label{A.23}
P_{i j}(\vec{\ell},\xi) = \sum_{r} \, P_{i j}^{r} (\vec{\ell}) \, \mathcal{L}_{r} (\xi),
\end{align}
where,
\begin{align}\label{A.24}
P_{i j}^{r} (\vec{\ell}) = \frac{2 r+1}{2} \, \int_{-1}^1 d\xi \, P_{i j}(\vec{\ell},\xi) \, \mathcal{L}_{r} (\xi).
\end{align}
We define the estimator for this multipole power spectrum in the following way,
\begin{eqnarray}\label{A.25}
\hat{P}_{i j}^{r} (\vec{\ell_a})&=&A_f \, \int_{\vec{\ell}_a} \frac{d^2 \vec{\ell}}{A_s (\vec{\ell}_a)} \, \frac{2 r+1}{2} \, 
 \int_{-1}^1 d\xi \, \kappa_i(\vec{\ell},\xi) 
\nonumber \\
&\times & \kappa_j(-\vec{\ell},\xi) \, \mathcal{L}_{r} (\xi),
\end{eqnarray}
being $\int_{\vec{\ell}_a} d^2 \vec{\ell} = A_s (\vec{\ell}_a) = \ell_a d\ell_a d\phi_a$, where $\cos \phi_a = \Upsilon_a$ (\ref{s1.72}). If we consider,
\begin{align}\label{A.26}
\langle \kappa_i(\vec{\ell}_1, \xi) \kappa_j(\vec{\ell}_2, \xi) \rangle = \delta_D (\vec{\ell}_1 + \vec{\ell}_2) \, P_{i j}(\vec{\ell}_1, \xi),
\end{align}
we can prove that $\langle \hat{P}_{i j}^{r} (\vec{\ell_a}) \rangle = P_{i j}^{r} (\vec{\ell_a})$. With this estimator we can calculate the covariance matrix,
\begin{align}\label{A.27}
C_{i j i' j'}^{r r'} (\vec{\ell}_a, \vec{\ell}_b) = \langle \hat{P}_{i j}^{r}(\vec{\ell}_a) \hat{P}_{i' j'}^{r'}(\vec{\ell}_b) \rangle - P_{i j}^{r}(\vec{\ell}_a) P_{i' j'}^{r'}(\vec{\ell}_b).
\end{align}
As in the previous case, we consider only gaussian perturbations, so that
\begin{align}\label{A.28}
C_{i j i' j'}^{r r'} (\vec{\ell}_a, \vec{\ell}_b) &= A_f^2 \, \frac{(2 r+1)(2 r'+1)}{4} \, \int_{\vec{\ell}_a} \frac{d^2 \vec{\ell}}{A_s (\vec{\ell}_a)} \, \int_{\vec{\ell}_b} \frac{d^2 \vec{\ell'}}{A_s (\vec{\ell}_b)} \nonumber\\
& \int_{-1}^1 d\xi \, \int_{-1}^1 d\xi' \mathcal{L}_{r} (\xi) \, \mathcal{L}_{r'} (\xi') \nonumber\\
& \left[ \langle \kappa_i(\vec{\ell}, \xi) \kappa_{i'}(\vec{\ell}', \xi') \rangle \, \langle \kappa_j(-\vec{\ell}, \xi) \kappa_{j'}(\vec{-\ell}', \xi') \rangle \right. \nonumber\\
& + \left. \langle \kappa_i(\vec{\ell}, \xi) \kappa_{j'}(-\vec{\ell}', \xi') \rangle \, \langle \kappa_j(-\vec{\ell}, \xi) \kappa_{i'}(\vec{\ell}', \xi') \rangle \right].
\end{align}
At this stage we use once again the distant observer approximation, in which we assume that the integrand of (\ref{A.28}) is non negligible only when $\xi' \simeq \xi$, then we obtain,
\begin{align}\label{A.29}
C_{i j i' j'}^{r r'} (\vec{\ell}_a, \vec{\ell}_b) &\simeq A_f^2 \, \frac{(2 r+1)(2 r'+1)}{2} \, \int_{\vec{\ell}_a} \frac{d^2 \vec{\ell}}{A_s (\vec{\ell}_a)} \, \int_{\vec{\ell}_b} \frac{d^2 \vec{\ell'}}{A_s (\vec{\ell}_b)} \nonumber\\
& \int_{-1}^1 d\xi \, \mathcal{L}_{r} (\xi) \, \mathcal{L}_{r'} (\xi) \nonumber\\
& \left[ \langle \kappa_i(\vec{\ell}, \xi) \kappa_{i'}(\vec{\ell}', \xi) \rangle \, \langle \kappa_j(-\vec{\ell}, \xi) \kappa_{j'}(\vec{-\ell}', \xi) \rangle \right. \nonumber\\
& + \left. \langle \kappa_i(\vec{\ell}, \xi) \kappa_{j'}(-\vec{\ell}', \xi) \rangle \, \langle \kappa_j(-\vec{\ell}, \xi) \kappa_{i'}(\vec{\ell}', \xi) \rangle \right].
\end{align}
Using (\ref{A.26}) and taking into account once more $\delta_D^2(x) = \delta_D (0) \delta_D (x)$ and $\delta_D (0) = 1/A_f$, we finally obtain,
\begin{align}\label{A.30}
C_{i j i' j'}^{r r'} (\vec{\ell}_a, \vec{\ell}_b) &\approx A_f \, \frac{(2 r+1)(2 r'+1)}{2} \, \int_{\vec{\ell}_a} \frac{d^2 \vec{\ell}}{A_s (\vec{\ell}_a)} \, \int_{\vec{\ell}_b} \frac{d^2 \vec{\ell'}}{A_s (\vec{\ell}_b)} \nonumber\\
& \int_{-1}^1 d\xi \, \mathcal{L}_{r} (\xi) \, \mathcal{L}_{r'} (\xi) \nonumber\\
& \left[ \delta_D(\vec{\ell}+\vec{\ell'}) P_{i i'}(\vec{\ell}, \xi) \, P_{j j'}(-\vec{\ell}, \xi)  \right. \nonumber\\
& + \left. \delta_D(\vec{\ell}-\vec{\ell'}) P_{i j'}(\vec{\ell}, \xi) \, P_{j i'}(-\vec{\ell}, \xi) \right],
\end{align}
As done before, we consider that $P_{i j}(\vec{\ell}, \xi) \simeq P_{i j}(\vec{\ell_a}, \xi)$ in the integral and also that,
\begin{align}\label{A.31}
\int_{\vec{\ell}_a} d^2 \vec{\ell} \, \int_{\vec{\ell}_b} d^2 \vec{\ell'} \,\, \delta_D(\vec{\ell} - \vec{\ell'}) = A_s (\ell_b) \, \delta_{\ell_a \ell_b} \, \delta_{\phi_a \phi_b},
\end{align}
and
\begin{align}\label{A.32}
\int_{\vec{\ell}_a} d^2 \vec{\ell} \, \int_{\vec{\ell}_b} d^2 \vec{\ell'} \,\, \delta_D(\vec{\ell} + \vec{\ell'}) = A_s (\ell_b) \, \delta_{\ell_a \ell_b} \, \delta_{\phi_a \phi_b+\pi}, 
\end{align}
so that we finally obtain,
\begin{widetext}
\begin{align}\label{A.33}
C_{i j i' j'}^{r r'} (\vec{\ell}_a, \vec{\ell}_b) =& \frac{A_f \, \delta_{\ell_a \ell_b}}{A_s (\vec{\ell}_a)} \, \frac{(2 r+1)(2 r'+1)}{2} \int_{-1}^1 d\xi \, \mathcal{L}_{r} (\xi) \, \mathcal{L}_{r'} (\xi) \nonumber \\
\times &\left[  \delta_{\phi_a \phi_b+\pi} \, P_{i i'}(\vec{\ell_a}, \xi) \, P_{j j'}(-\vec{\ell}_a, \xi) + \delta_{\phi_a \phi_b} \, P_{i j'}(\vec{\ell_a}, \xi) \, P_{j i'}(-\vec{\ell}_a, \xi) \right].
\end{align}
\end{widetext}
Notice that if the only dependence in $\phi_a$ is in the form of $\Upsilon_a^2$, then $\delta_{\phi_a \phi_b} = \delta_{\phi_a \phi_b+\pi} = \delta_{\Upsilon_a \Upsilon_b}$. If we further consider that the fiducial power spectrum is isotropic, we obtain the final result used in the work,
\begin{widetext}
\begin{align}\label{A.34}
&C_{i j i' j'}^{r r'} (\vec{\ell}_a, \vec{\ell}_b) = \frac{2 \pi \, (2r+1) \, \delta_{\vec{\ell}_a \vec{\ell}_b} \delta_{r r'}}{f_{sky} \,(2\ell_a + 1) d\ell_a d\phi_a} \, [P^{obs}_{i i'}(\ell_a) P^{obs}_{j j'}(\ell_a) + P^{obs}_{i j'}(\ell_a) P^{obs}_{j i'}(\ell_a)],
\end{align}
\end{widetext}
where $\delta_{\vec{\ell}_a \vec{\ell}_b} = \delta_{\ell_a \ell_b} \delta_{\Upsilon_a \Upsilon_b}$ and we have approximated once more $\ell\simeq (2\ell + 1)/2$.

\end{document}